\documentclass[english]{article}
\usepackage[T1]{fontenc}
\usepackage[latin9]{inputenc}
\usepackage{textcomp}
\usepackage{amsmath}
\usepackage{graphicx}
\usepackage{amssymb}
\usepackage{esint}

\makeatletter
\newcommand{\lyxaddress}[1]{
\par {\raggedright #1
\vspace{1.4em}
\noindent\par}
}

\@ifundefined{definecolor}
 {\usepackage{color}}{}
\makeatother

\makeatother

\usepackage{babel}

\makeatother

\usepackage{babel}

\makeatother

\usepackage{babel}

\makeatother

\usepackage{babel}

\makeatother

\usepackage{babel}

\makeatother

\usepackage{babel}

\makeatother

\usepackage{babel}

\begin{document}

\title{On the Dynamics of the Fermi-Bose Model}

\author{M \"{O}gren$^{1}$ and M Carlsson$^{2}$}

\date{\today{}}

\maketitle

\lyxaddress{$^{1}$Department of Mathematics, Technical University of Denmark,
2800 Kgs. Lyngby, Denmark.\\
 $^{2}$Center for Mathematical Sciences, Lund University, Box
118, 22100 Lund, Sweden.}
\begin{abstract}
We consider the exponential matrix representing the
dynamics of the Fermi-Bose model in an undepleted bosonic field approximation.
A recent application of this model is molecular dimers dissociating
into its atomic compounds. The problem is solved in $D$ spatial dimensions
by dividing the system matrix into blocks with generalizations of
Hankel matrices, here refered to as $D$-block-Hankel matrices. The
method is practically useful for treating large systems, i.e. dense
computational grids or higher spatial dimensions, either on a single
standard computer or a cluster. In particular the results can be used
for studies of three-dimensional physical systems of arbitrary geometry.
We illustrate the generality of our approach by giving numerical results
for the dynamics of Glauber type atomic pair correlation functions
for a non-isotropic three-dimensional harmonically trapped molecular
Bose-Einstein condensate. 
\end{abstract}

\section{Introduction}

The Fermi-Bose model under study here forms the underlying basis for
a range of phenomena in condensed matter and ultra-cold atomic physics.
It was proposed in the context of high-temperature superconductivity
by Friedberg and Lee \cite{FriedbergPRB1989}, and in ultracold gases
it corresponds to the theory of resonance superfluidity with Feshbach
molecules \cite{KheruntsyanPRA2000,HollandPRL2001,TimmermansPLA2001}.
The latter forms the basis of a model for describing the physics of
the BCS-BEC crossover \cite{OhashiPRL2002}. More recently, the fermion-boson
model has been used for analyzing the decay of double occupancies
(doublons) \cite{StrohmaierPRL2010} in a driven Fermi-Hubbard system
\cite{JordensNature2008}. The particular situation that we concentrate
on in this article corresponds to spontaneous dissociation of a Bose-Einstein
condensate of molecular dimers into fermionic atoms \cite{JackPRA2005,Fermidiss}.
This process represents a fermionic counterpart of parametric down-conversion
in quantum optics. After recent experimental achivements of molecular
dissociation \cite{MukaiyamaPRL2004,DurrPRA2004,ThompsonPRL2005,GreinerPRL2005},
our aim here is to take the theory for numerical modeling of dissociation
into fermionic atoms from the state of principally possible to the
state of being useful in practice.

\subsection{Effective quantum field theory }

Let us here in brief present the three-wave interaction type Hamiltonian
of interest in this work \cite{LeePR1954,Kaellen1955,Heisenberg1957}.

\begin{equation}
\hat{H}=\hat{H}_{0}-i\hbar\chi\int d\mathbf{x}\left(\hat{\Psi}_{0}^{\dagger}\hat{\Psi}_{1}\hat{\Psi}_{2}-\hat{\Psi}_{2}^{\dagger}\hat{\Psi}_{1}^{\dagger}\hat{\Psi}_{0}\right).\label{eq:HinSpatialCoordinates}\end{equation}
 Here all the quadratic terms are collected in $\hat{H}_{0}$ and
contains kinetic- and potential energy terms, $\hat{\Psi}_{0}(\mathbf{x},t)$
stands for a bosonic field operator, whereas $\hat{\Psi}_{j}(\mathbf{x},t)$
($j=1,2$) describe two particle fields that can be two fermions (bosons)
in different spin states, finally $\chi$ is the strength of the fermion-boson
(boson-boson) coupling term.

\subsection{Applications of the Fermi-Bose model}

In modern condensed matter physics the Fermi-Bose model have two major
areas of applicability. First the so called {}``s-channel'' model
in high-temperature superconductivity \cite{FriedbergPRB1989}. In
this context it model the formation dynamics of bosonic Cooper-pairs,

\begin{picture}(400,60) 

\put(185,32){\vector(1,0){60}} \put(135,32){\vector(-1,0){60}}

\put(245,28){\vector(-1,0){60}} \put(75,28){\vector(1,0){60}}

\put(60,30){\circle*{20}} \put(260,30){\circle{15}}

\put(149,30){\circle*{20}} \put(171,30){\circle{15}}

\put(157,24){\textasciitilde{}}

\put(60,13){\vector(0,1){34}} \put(260,47){\vector(0,-1){34}}

\put(149,13){\vector(0,1){34}} \put(171,47){\vector(0,-1){34}}

\put(130,1){\textsc{Cooper-pair}} 
\put(55,1){\textsc{$e_{\uparrow}^{-}$}}

\put(255,1){\textsc{$e_{\downarrow}^{-}$}}

\end{picture}

\vspace{2mm}
 where the two atomic particles, the electrons, are fermions.

\noindent In the field of ultra-cold atomic physics, it can model
the dissociation of ultra-cold bosonic molecules \cite{PoulsenPRA2001,DrummondPRA2002},

\begin{picture}(400,60) 

\put(180,32){\vector(1,0){65}} \put(140,32){\vector(-1,0){65}}

\put(245,28){\vector(-1,0){65}} \put(75,28){\vector(1,0){65}}

\put(60,30){\circle*{20}} \put(260,30){\circle{15}}

\put(155,30){\circle*{20}} \put(165,30){\circle{15}}

\put(160,30){\circle{25}}

\put(60,13){\vector(0,1){34}} \put(260,47){\vector(0,-1){34}}

\put(155,13){\vector(0,1){34}} \put(165,47){\vector(0,-1){34}}

\put(135,1){\textsc{molecule}} \put(42,1){\textsc{atom 1}} \put(242,1){\textsc{atom
2}}

\end{picture}

\vspace{2mm}
and hence we allow here for the atomic particles to be either two
fermions or two bosons.

\subsection{Computational methods for large systems}

\noindent Various generalizations of time-dependent DMRG \cite{CazalillaPRL2002}
to higher dimensional systems is a topic of large present interest
in the computational physics community \cite{Schollwock2011}, but
have not yet reached a useful status for large higher dimensional
systems as of interest here.

Methods for bosonic evolution based on stochastic differential equations
(SDEs), with the ability for independent stochastic trajectories to
be carried out on computer clusters for large systems, are succesful
in many situations \cite{PoulsenPRA2001,DrummondPRA2002,DeuarPRA2011,KrachmalnicoffPRL2010,OgrenPRA2010,SavagePRA2006,MidgleyPRA2009}
but are restricted to simulations of molecules dissociating into bosonic
atoms.

In this article we outline a method to study the fermionic time evolution
for effective Heisenberg equations that are linear in creation and
annihilation operators. We apply the method to the evaluation of analytic
short-time asymptots for the Glauber's second order correlation functions
\cite{OgrenPRA2010}, for a non-isotropic three-dimensional molecular
condensate dissociating into fermionic atoms, against numerical data.
We focus first on a general formulation in a $D$-dimensional Cartesian
momentum base, such that convenient numerical solutions can be directly
obtained for an arbitrary shaped bosonic field. However, for a specific
application the numerical performance may be further increased by
using additional geometrical symmetries or by formulating the equations
in a different basis. Except for generating valuable results in certain
physical regimes, the method presented here can also be useful as
a reference for validation of approximate analytic results or to evaluate
more advanced numerical methods, such as for example the Gaussian
fermionic phase-space representation (GPSR) that have recently been
applied to the Fermi-Bose model with a uniform bosonic field \cite{Magnus-Karen-Joel-First-Principles,OgrenCPC2010}
and an implementation of GPSR for dissociation from a non-uniform
molecular BEC is in progress. Turning off stochastic terms in the
SDEs of the GPSR, one obtain in effect the so called pairing mean-field
theory (PMFT) \cite{JackPRA2005,Magnus-Karen-Joel-First-Principles,PMFT}.
Furthermore, keeping the molecular variables in PMFT undepleted will
be equivalent to the formulation in the present article but is numerically
less suitable.\textbf{ }

There is an obvious computational advantage to solve for the single
operator dynamics, instead of solving for pairs of operators, the
latter is done e.g. in the PMFT (and GPSR) discussed above. To give
one relevant example, to simulate a three-dimensional non-uniform
field on a Cartesian momentum grid of size $100\times100\times100$
requires in effect, as we will show in this paper, only to be able
to store part of a (sparse) $n\times n$ $D$-block-Hankel matrix
of the size $n=100^{3}=10^{6}$ and to multiply this $D$-block-Hankel
matrix with vectors only for each final time-point of interest. On
the other hand c-number based mean-field methods for Fermi-Bose systems
like PMFT, where the basic variables represents pairs of operators
with two indices, requires in this case to propagate $\sim10^{12}$
variables in time through sufficiently many small time-steps up to
the final time-point of interest. The methods for stochastic evolution
that represents single bosonic field operators with complex stochastic
fields mentioned above, see \cite{DeuarPRA2011} for a recent review,
do not have any direct corresponding useful method for fermions. For
example the GPSR that can treat the quantum dynamics of the Fermi-Bose
model exact \cite{Magnus-Karen-Joel-First-Principles,OgrenCPC2010,CorneyPRL2004},
involves basis elements that represents pairs of single operators
\cite{Magnus-Karen-Joel-First-Principles,OgrenCPC2010,CorneyPRL2004,CorneyJoPA2006,RahavPRB2009,RosalesPRA2011},
hence it is generally more restrictive in the size of the computational
grid.

Due to the existence of several related methods for bosonic dynamics,
we focus in the present article in particular on applications to systems
with fermionic atomic operators. However, the specific analogue theory
of two distinguishable bosonic atomic operators is also presented
for comparison, since its formulation differ only with a sign.

Advantages with the method presented here are \emph{i)} to be able
to study effects of quantum statistics {[}by chosing $q=-1$ ($q=1$)
for fermionic (bosonic) atomic particles{]}; \emph{ii)} to obtain
accurate correlation functions for short times i.e. for a small number
of atomic particles, where stochastic evolution methods have a low
signal to noise ratio; and \emph{iii)} to conviniently treat systems
with a moderate size (relative to the RAM memory) of the computational
grid (e.g. lower dimensional systems), where deterministic values
of the observables are obtained fast even on a single standard PC.

The article is organized as follows. Section~\ref{sec:The-momentum-space-operator}
provides the Heisenberg equation of motion to be treated within the
concept of molecular dissociation into fermionic (bosonic) atoms and
also briefly mention dissociation into two indistinguishable bosonic
atoms and the related problem of condensate collision. In section~\ref{sec:Non-uniform-molecular-field}
we relate blocks in the system matrix responsible for atom-molecule
coupling to the so called $D$-block-Hankel matrices. In section~\ref{sub:A-Theorem-for}
we formulate and prove results for the structure of the solution of
Heisenberg equation of motion. Section~\ref{sub:Obtaining-the-observables}
provides the reader with the practical details of how to use the mathematical
results in obtaining physical observables. We apply the general method
to a three-dimensional non-isotropic harmonically trapped Bose-Einstein
condensate in section~\ref{sec:Results} and present numerical results
for atomic correlation functions that we compare to recently derived
analytic short-time asymptotes. Finally the article is summarised
in section~\ref{sec:Summary}.

\section{The momentum-space operator equations\label{sec:The-momentum-space-operator}}

\label{sec:Model-dist}

To model the dissociations of a Bose-Einstein condensate of diatomic
molecules into pairs of constituent atoms, we start with the following
effective quantum field theory Hamiltonian

\[
\widehat{H}=\int d\mathbf{x}\biggl\{\sum\limits _{j=0,1,2}\frac{\hbar^{2}}{2m_{j}}|\nabla\widehat{\Psi}_{j}|^{2}+\hbar\Omega(\widehat{\Psi}_{1}^{\dagger}\widehat{\Psi}_{1}+\widehat{\Psi}_{2}^{\dagger}\widehat{\Psi}_{2})\]

\begin{equation}
-i\hbar\chi\left(\widehat{\Psi}_{0}^{\dagger}\widehat{\Psi}_{1}\widehat{\Psi}_{2}-\widehat{\Psi}_{2}^{\dagger}\widehat{\Psi}_{1}^{\dagger}\widehat{\Psi}_{0}\right)\biggr\}.\label{hamiltonian}\end{equation}
 Here we assume that the molecules are made of either two distinguishable
bosonic atoms or two fermionic atoms in different spin states. In
both cases, $\widehat{\Psi}_{0}(\mathbf{x},t)$ is a bosonic field
operator for the molecules, satisfying the standard commutation relations
$[\widehat{\Psi}_{0}(\mathbf{x},t),\widehat{\Psi}_{0}^{\dagger}(\mathbf{x}^{\prime},t)]=\delta^{D}(\mathbf{x}-\mathbf{x}^{\prime})$,
with $D$ being the spatial dimension of the system \cite{FetterWalecka}.
The atomic field operators, $\widehat{\Psi}_{j}(\mathbf{x},t)$ ($j=1,2$),
satisfy either bosonic commutation or fermionic anti-commutation relations,
$[\widehat{\Psi}_{i}(\mathbf{x},t),\widehat{\Psi}_{j}^{\dagger}(\mathbf{x}^{\prime},t)]=\delta_{ij}\delta^{D}(\mathbf{x}-\mathbf{x}^{\prime})$
and $[\widehat{\Psi}_{i}^{\dagger}(\mathbf{x},t),\widehat{\Psi}_{j}^{\dagger}(\mathbf{x}^{\prime},t)]=[\widehat{\Psi}_{i}(\mathbf{x},t),\widehat{\Psi}_{j}(\mathbf{x}^{\prime},t)]=0$
or $\{\widehat{\Psi}_{i}(\mathbf{x},t),\widehat{\Psi}_{j}^{\dagger}(\mathbf{x}^{\prime},t)\}=\delta_{ij}\delta^{D}(\mathbf{x}-\mathbf{x}^{\prime})$
and $\{\widehat{\Psi}_{i}^{\dagger}(\mathbf{x},t),\widehat{\Psi}_{j}^{\dagger}(\mathbf{x}^{\prime},t)\}=\{\widehat{\Psi}_{i}(\mathbf{x},t),\widehat{\Psi}_{j}(\mathbf{x}^{\prime},t)\}=0$,
depending on the underlying statistics \cite{FetterWalecka}.

The first term in the Hamiltonian (\ref{hamiltonian}) describes the
kinetic energy where the atomic masses are $m_{1}$ and $m_{2}$,
whereas the molecular mass is $m_{0}=m_{1}+m_{2}$. For simplicity,
we will consider the case of equal atomic masses (same isotope atoms),
with $m_{1}=m_{2}\equiv m_{a}$ and $m_{0}=2m_{a}$.

The coupling constant $\chi\equiv\chi_{D}$ is responsible for coherent
conversion of molecules into atom pairs, e.g. via optical Raman transitions,
an rf transition, or a Feshbach resonance sweep and microscopic expressions
for $\chi$ can be found in \cite{PMFT} and references therein. The
detuning $\Omega$ is defined to give the overall energy mismatch
$2\hbar\Omega$ between the free two-atom state in the dissociation
threshold and the bound molecular state (including the relative frequencies
of the Raman lasers or the frequency of the rf field, again see \cite{PMFT}
and references therein for details). Unstable molecules, spontaneously
dissociating into pairs of constituent atoms, correspond to $\Omega<0$,
with $2\hbar|\Omega|$ being the total dissociation energy.

In what follows we will treat the dissociation dynamics in the undepleted
molecular condensate approximation in which the molecules are represented
as a fixed classical field. The approximation is valid for short enough
dissociation times during which the converted fraction of molecules
does not exceed about $10\%$ \cite{SavagePRA2006,PMFT,Ogren-directionality}.
In this regime the dissociation typically produces low density atomic
clouds for which the atom-atom $s$-wave scattering interactions are
negligible. For dissociation into bosonic atoms, also the effects
of atom-atom $s$-wave scattering have been investigated in \cite{OgrenPRA2010,SavagePRA2006,MidgleyPRA2009},
and the validity of the undepleted molecular condensate approximation
was found to still hold for a converted fraction of molecules up to
about 5\%. Even though 5\%\textminus{}10\% conversion efficiencies
seem small, nevertheless they can produce mesoscopic ensembles of
pair-correlated atoms with interesting quantum statistics and nontrivial
many-body correlations if one starts with large-enough molecular condensates,
such as containing at least $10^{4}$\textminus{}$10^{5}$ molecules.
Further on, for dissociation into fermionic atoms in the regime of
Pauli-blocking, where the atomic occupation numbers are strictly limited
by the number of available states \cite{PMFT}, the undepleted field
approximation is expected to be accurate even for large times \cite{Fermidiss,Magnus-Karen-Joel-First-Principles,PMFT}.
In addition to the issue about temporal depletion, the atom-molecule
interactions will initially appear as an effective spatially dependent
detuning due to the mean-field interaction energy~\cite{DeuarEPJD2011}. 
To neglect this
effect can be motivated by operating at relatively large absolute
values of the dissociation detuning $|\Omega|$ so that it dominates
the mean-field energy shift \cite{SavagePRA2006,Ogren-directionality}.

The trapping potential for preparing the initial molecular BEC --
with any residual atoms being removed -- is omitted from the Hamiltonian
(\ref{hamiltonian}) since we assume that once the dissociation is
invoked, the trapping potential is switched off, so that the dynamics
of dissociation is taking place in free $D$-dimensional space. We
assume that the switching on of the atom-molecule coupling $\chi$
and switching off of the trapping potential is done in a sudden jump
at time zero. Accordingly the preparation stage is reduced to assuming
a certain initial state of the molecular BEC in a trap, after which
the dynamics is governed by the Hamiltonian (\ref{hamiltonian}).

\subsection{Heisenberg equations in the undepleted molecular condensate approximation}

From the Heisenberg equation of motion with the Hamiltonian taken
from (\ref{hamiltonian}), we have for the three field operators \begin{equation}
\frac{\partial\hat{\Psi}_{j}\left(\mathbf{x}\right)}{\partial t}=-\frac{i}{\hbar}\left[\hat{\Psi}_{j}\left(\mathbf{x}\right),\,\hat{H}\right],\ j=0,\:1,\:2.\label{eq:Heisenberg_Eq_for_j_1_2_3}\end{equation}

In order to obtain linear operator equations the undepleted molecular
field approximation is first invoked as follows. Assuming that the
molecules are in a coherent state initially, the density profile $\rho_{0}\left(\mathbf{x}\right)$
is in principal given by the ground state solution of the standard
Gross-Pitaevskii equation. We then replace the molecular field operator
by its coherent mean-field complex function \cite{Fermidiss,OgrenPRA2010},
the so called condensate wave-function, \begin{equation}
\widehat{\Psi}_{0}(\mathbf{x},t)\rightarrow\langle\widehat{\Psi}_{0}(\mathbf{x},t)\rangle\equiv\Psi_{0}(\mathbf{x},0)=\sqrt{\rho_{0}\left(\mathbf{x}\right)}\exp\left(i\theta\left(\mathbf{x}\right)\right).\label{apa}\end{equation}
 From (\ref{hamiltonian}), (\ref{eq:Heisenberg_Eq_for_j_1_2_3})
and (\ref{apa}) we then write down the Heisenberg equations for the
remaining two coupled atomic field operators as follows

\begin{equation}
\frac{\partial\widehat{\Psi}_{1}\left(\mathbf{x},t\right)}{\partial t}=i\left[\frac{\hbar}{2m_{a}}\nabla^{2}-\Omega\right]\widehat{\Psi}_{1}\left(\mathbf{x},t\right)+q\chi\sqrt{\rho_{0}\left(\mathbf{x}\right)}\exp\left(i\theta\left(\mathbf{x}\right)\right)\widehat{\Psi}_{2}^{\dag}\left(\mathbf{x},t\right),\label{eq:Heisenberg-eqs}\end{equation}

\begin{equation}
\frac{\partial\widehat{\Psi}_{2}^{\dagger}\left(\mathbf{x},t\right)}{\partial t}=-i\left[\frac{\hbar}{2m_{a}}\nabla^{2}-\Omega\right]\widehat{\Psi}_{2}^{\dagger}\left(\mathbf{x},t\right)+\chi\sqrt{\rho_{0}\left(\mathbf{x}\right)}\exp\left(-i\theta\left(\mathbf{x}\right)\right)\widehat{\Psi}_{1}\left(\mathbf{x},t\right).\label{eq:Heisenberg-eqs_lower}\end{equation}
 The sign given by $q$ in the second term in (\ref{eq:Heisenberg-eqs})
is $q=-1$ for fermionic and $q=1$ for bosonic atoms throughout the
paper, as a consequence of different operator (anti-) commutator relations.
Multiplying (\ref{eq:Heisenberg-eqs}) and (\ref{eq:Heisenberg-eqs_lower})
with $L^{-D/2}\exp(-i\mathbf{k\cdot x})$, where $V=L^{D}$ is the
quantization volume and we assume for simplicity $L$ to be the spatial
length of the system in any direction, followed by integration over
$\mathbf{x}$, we can interpret the differential equations for a given
$\mathbf{k}$ in terms of the Fourier operators in momentum space

\begin{equation}
\widehat{a}_{\mathbf{k},j}(t)=\frac{1}{L^{D/2}}\int_{V}d\mathbf{x}\,\widehat{\Psi}_{j}(\mathbf{x},t)\exp(-i\mathbf{k\cdot x}).\label{def:FouirerOperator}\end{equation}
 The operators $\widehat{a}_{\mathbf{k},j}(t)$ satisfy the usual
(commutation-) anti-commutation relations $[\widehat{a}_{\mathbf{k},i},\widehat{a}_{\mathbf{k}',j}^{\dagger}]_{-q}=\delta_{ij}\delta_{\mathbf{k},\mathbf{k}^{\prime}}$
and $[\widehat{a}_{\mathbf{k},i}^{\dagger},\widehat{a}_{\mathbf{k}',j}^{\dagger}]_{-q}=[\widehat{a}_{\mathbf{k},i},\widehat{a}_{\mathbf{k}',j}]_{-q}=0$
(i.e. for $q=-1$, $\left[\:,\:\right]_{+1}\equiv\left\{ \:,\:\right\} $).
Since the effective Hamiltonian corresponding to (\ref{eq:Heisenberg-eqs})-(\ref{eq:Heisenberg-eqs_lower})
is quadratic in the field operators, higher-order moments or expectation
values of products of creation and annihilation operators will factorize
according to Wick's theorem into products of the normal and anomalous
densities $n_{\mathbf{k},\mathbf{k}^{\prime},j}\equiv$$\left\langle \widehat{a}_{\mathbf{k},j}^{\dagger}\widehat{a}_{\mathbf{k}',j}\right\rangle $
and $m_{\mathbf{k},\mathbf{k}^{\prime}}\equiv\left\langle \widehat{a}_{\mathbf{k},1}\widehat{a}_{\mathbf{k}',2}\right\rangle $.
Upon applying the Fourier transform, the equations (\ref{eq:Heisenberg-eqs})
and (\ref{eq:Heisenberg-eqs_lower}) become

\begin{equation}
\frac{d\widehat{a}_{\mathbf{k},1}}{dt}=-i\Delta_{\mathbf{k}}\widehat{a}_{\mathbf{k},1}+q\kappa\sum_{\mathbf{k}'}\tilde{g}_{\mathbf{k}'+\mathbf{k}}\widehat{a}_{\mathbf{k}',2}^{\dagger},\label{eq:MatrixAppHE_first_eq}\end{equation}

\begin{equation}
\frac{d\widehat{a}_{\mathbf{k},2}^{\dagger}}{dt}=\kappa\sum_{\mathbf{k}'}\tilde{g}_{\mathbf{k}'+\mathbf{k}}^{*}\widehat{a}_{\mathbf{k}',1}+i\Delta_{\mathbf{k}}\widehat{a}_{\mathbf{k},2}^{\dagger}.\;\;\;\label{eq:MatrixAppHE}\end{equation}
 where $*$ denotes the complex conjugate. The kinetic part is $\Delta_{\mathbf{k}}\equiv\Omega+\hbar\left|\mathbf{k}\right|^{2}/\left(2m_{a}\right)$.
The effective atom-molecule coupling constant is $\kappa=\chi/L^{D/2}$
\cite{PMFT}. Finally the complex Fourier coefficients $\tilde{g}_{\mathbf{k}}$
of the condensate wave function describing the molecular mean-field
in momentum-space is defined analogue to (\ref{def:FouirerOperator})

\begin{equation}
\tilde{g}_{\mathbf{k}}=\frac{1}{L^{D/2}}\int_{V}d\mathbf{x}\sqrt{\rho_{0}\left(\mathbf{x}\right)}\exp\left(i\theta\left(\mathbf{x}\right)-i\mathbf{k}\cdot\mathbf{x}\right),\label{defFourierCoefficients}\end{equation}
For a real condensate wavefunction we have $\tilde{g}_{\mathbf{k}'+\mathbf{k}}^{*}=\tilde{g}_{-\left(\mathbf{k}'+\mathbf{k}\right)}$,
while a non-zero phase-function $\theta\left(\mathbf{x}\right)$ can
be used to represent for example an initial vortex state as in \cite{PoulsenPRA2007}.
We give the general theory for a static complex condensate wave-function
$\Psi_{0}=\sqrt{\rho_{0}\left(\mathbf{x}\right)}\exp\left(i\theta\left(\mathbf{x}\right)\right)$
in the following, while we for the numerical example in section \ref{sec:Results}
choose $\Psi_{0}=\sqrt{\rho_{0}\left(\mathbf{x}\right)}$ to be real.
For the case where the molecular density is uniform and constant $\rho_{0}$,
there is only one non-zero Fourier coefficient and only operators
with index $\mathbf{k}$ and $\mathbf{k}'=-\mathbf{k}$ couples in
(\ref{eq:MatrixAppHE_first_eq})-(\ref{eq:MatrixAppHE}). Finally,
if the molecular density is non-uniform and time dependent $\Psi\left(\mathbf{x},t\right)=\sqrt{\rho\left(\mathbf{x},t\right)}\exp\left(i\theta\left(\mathbf{x},t\right)\right)$,
we need to determine the Fourier coefficients (\ref{defFourierCoefficients})
for each time-step in an iterative process where the dynamics of the
molecular mean-field is taken into account. This is a topic for future
work, however, the major effects to be taken into account are: expansion
of the condensate \cite{CastinPRL1996}; temporal depletion of the
number of molecules \cite{OgrenPRA2010,SavagePRA2006,MidgleyPRA2009,Magnus-Karen-Joel-First-Principles,PMFT};
and finally the spatial dependences, such as a larger local dissociation
rate in the points of highest initial molecular density \cite{MidgleyPRA2009}.
Defining such an iterative process, the results of the present article
will be applicable in each time-step of the evolution.

When the molecular condensate consists of pairs of indistinguishable
bosonic atoms of a single spin-state, for example $^{87}$Rb$_{2}$
\cite{DurrPRA2004}, Heisenberg equation for one bosonic operator
describe the dissociation dynamics \cite{OgrenPRA2010,PMFT}. A similar
set of Heisenberg equations for bosonic operators of one spin-state
can also be formulated for the dynamics of condensate collisions within
the time-dependent Bogoliubov approach \cite{DeuarPRA2011,KrachmalnicoffPRL2010,Trippenbach,Magnus-Karen-4WM}.
Hence, also for this problem the method discussed in the present article
can be applied \cite{Ogrenunpublished2007}.

\subsection{Uniform molecular field\label{sub:Uniform--dimensional-MBEC}}

For a reference we start with presenting the relevant results for
a size-matched uniform system. Size-matched here means that the spatial
dimensions of the uniform bosonic field $L_{eq}$, with the same (central-)
particle density $\rho_{0}$, are chosen such that the initial number
of molecules are the same as for the non-uniform system of interest,
hence $N_{mol}=\rho_{0}L_{eq}^{D}$.

Under the condition of a uniform molecular field, i.e. that do not
depend on the spatial coordinates $\Psi_{0}\left(\mathbf{x}\right)=\Psi_{0}=\sqrt{\rho_{0}}$,
equations (\ref{eq:MatrixAppHE_first_eq})-(\ref{eq:MatrixAppHE})
with initial vacuum states for the atoms have analytical solutions
for the normal- and anomalous atomic moments $n_{\mathbf{k},\sigma}\equiv\left\langle \widehat{n}_{\mathbf{k},\sigma}\right\rangle \equiv\left\langle \widehat{a}_{\mathbf{k},1}^{\dagger}\widehat{a}_{\mathbf{k},1}\right\rangle =\left\langle \widehat{a}_{-\mathbf{k},1}^{\dagger}\widehat{a}_{-\mathbf{k},1}\right\rangle =\left\langle \widehat{a}_{\mathbf{k},2}^{\dagger}\widehat{a}_{\mathbf{k},2}\right\rangle =\left\langle \widehat{a}_{-\mathbf{k},2}^{\dagger}\widehat{a}_{-\mathbf{k},2}\right\rangle $
respectively $m_{\mathbf{k}}\equiv\left\langle \widehat{a}_{\mathbf{k},1}\widehat{a}_{-\mathbf{k},2}\right\rangle =\left\langle \widehat{a}_{-\mathbf{k},1}\widehat{a}_{\mathbf{k},2}\right\rangle $.
The related (PMFT) complex differential equations with initial conditions
$n_{\mathbf{k},\sigma}\left(0\right)=m_{\mathbf{k}}\left(0\right)=0$
are \cite{PMFT}

\begin{equation}
\frac{dn_{\mathbf{k},\sigma}}{dt}=2g_{0}\textnormal{{Re}}\left\{ m_{\mathbf{k}}\right\} ,\label{eq:PMFTeqsys1}\end{equation}

\begin{equation}
\frac{dm_{\mathbf{k}}}{dt}=-2i\Delta_{\mathbf{k}}m_{\mathbf{k}}+g_{0}\left(1+q2n_{\mathbf{k},\sigma}\right),\label{eq:PMFTeqsys2}\end{equation}
 where $g_{0}\equiv\kappa\tilde{g}_{0}=\chi\sqrt{\rho_{0}}$. The
corresponding solutions to (\ref{eq:PMFTeqsys1})-(\ref{eq:PMFTeqsys2}),
calculated explicitly in section \ref{sub:Comparison-with-the}, are

\begin{equation}
n_{\mathbf{k},\sigma}=\frac{g_{0}^{2}}{\Delta_{\mathbf{k}}^{2}-qg_{0}^{2}}\sin^{2}\left(\sqrt{\Delta_{\mathbf{k}}^{2}-qg_{0}^{2}}\, t\right),\label{eq:Uniform_nk}\end{equation}

\[
m_{\mathbf{k}}=\frac{g_{0}}{\sqrt{\Delta_{\mathbf{k}}^{2}-qg_{0}^{2}}}\cos\left(\sqrt{\Delta_{\mathbf{k}}^{2}-qg_{0}^{2}}\, t\right)\sin\left(\sqrt{\Delta_{\mathbf{k}}^{2}-qg_{0}^{2}}\, t\right)\]

\begin{equation}
\qquad-i\frac{g_{0}\Delta_{\mathbf{k}}}{\Delta_{\mathbf{k}}^{2}-qg_{0}^{2}}\sin^{2}\left(\sqrt{\Delta_{\mathbf{k}}^{2}-qg_{0}^{2}}\, t\right).\label{eq:Uniform_mk}\end{equation}
 Note that for bosons ($q=1$), e.g., the resonance mode ($\Delta_{\mathbf{k}}\equiv0$)
leads to a Bose-enhancement effect in the atomic occupations, described
by $n_{\mathbf{k}_{0},\sigma}\left(t\right)=\sinh^{2}\left(g_{0}t\right)$
which grows exponentially with time, consequently this illustrate
that the undepleted molecular field approximation (\ref{apa}) is
only realistic for a short time. In contrast to this, for fermionic
atoms, the atomic occupations undergo sinusoidal oscillations and
can be kept to a small fraction of the number of molecules also for
large times. As noted in \cite{Fermidiss} the moments (\ref{eq:Uniform_nk})
and (\ref{eq:Uniform_mk}) fulfills the equality

\begin{equation}
\left|m_{\mathbf{k}}\right|^{2}=n_{\mathbf{k},\sigma}\left(1+qn_{\mathbf{k},\sigma}\right),\label{eq:MatrixAppUniformSolutions}\end{equation}
 for a uniform molecular field.

\section{Non-uniform molecular field\label{sec:Non-uniform-molecular-field}}

We here give the theoretical framework needed for the main analytic
results of the article, which is given in the next section, for how
to efficiently solve the Fermi-Bose model for a non-uniform molecular
field. In particular we show how the system matrix for the dynamics
of the Fermi-Bose model of a physical system in $D$ spatial dimensions
can be classified in terms of generalizations of Hankel matrices.

In cases where the shape of the molecular condensate $\rho\left(\mathbf{x}\right)$
posseses certain geometrical symmetrices, such as spherical symmetry,
a reduction of the number of atomic creation and annihilation operators
can be implemented. Alternatively, a base with atomic operators that
are directly defined e.g. in a spherical coordinate system can be
used \cite{Trippenbach}. In this article, however, we treat the case
of a general dense cubic lattice. In practice, the set of indices
$\mathbf{k}$ (and $\mathbf{k'}$) lie on a finite grid in $\mathbb{R}^{D}$
of the form $\mathbf{k}=(k_{1},\ldots,k_{D})=\frac{2\pi}{L}\mathbf{n}$,
where $\mathbf{n}\in\mathbb{Z}^{D}$ and $-K\leq n_{j}\leq K$ for
all $j=1,\ldots,D$. By abuse of notation, we will often suppress
the factor $\frac{2\pi}{L}$ and identify $\mathbf{k}$ with $\mathbf{n}$,
i.e. we will write e.g. $\hat{a}_{\mathbf{n},1}$ in place of $\hat{a}_{\mathbf{k},1}$
wherever convenient. Whenever $\mathbf{n}$ or $\mathbf{n'}$ appears
it will be implicitly understood that it stays within the above limitations.

\subsection{One-dimensional systems}

To set the scene, we first discuss a system in one spatial dimension.
For a non-uniform system with $D=1$, set $B=2K+1$ and identify the
systems of annihilation operators $\{\hat{a}_{n,1}\}_{-K\leq n\leq K}$
and creation operators $\{\hat{a}_{n,2}^{\dagger}\}_{-K\leq n\leq K}$
with the $2B$ dimensional column vector $\left[\widehat{a}_{-K,1}\ldots\widehat{a}_{K,1}\widehat{a}_{-K,2}^{\dagger}\ldots\widehat{a}_{K,2}^{\dagger}\right]^{T}$.
Under this identification, Heisenberg equations (\ref{eq:MatrixAppHE_first_eq})-(\ref{eq:MatrixAppHE})
can then be visualized in terms of a $2B\times2B$-system-matrix $A$
composed of four blocks of $B\times B$-matrices

\begin{equation}
\frac{d}{dt}\left[\begin{array}{l}
\widehat{a}_{-K,1}\\
\vdots\\
\widehat{a}_{K,1}\\
\widehat{a}_{-K,2}^{\dagger}\\
\vdots\\
\widehat{a}_{K,2}^{\dagger}\end{array}\right]=\left[\begin{array}{cc}
A_{11} & A_{12}\\
A_{21} & A_{22}\end{array}\right]\left[\begin{array}{l}
\widehat{a}_{-K,1}\\
\vdots\\
\widehat{a}_{K,1}\\
\widehat{a}_{-K,2}^{\dagger}\\
\vdots\\
\widehat{a}_{K,2}^{\dagger}\end{array}\right].\label{ODEinMatrixForm}\end{equation}
 It then follows directly from (\ref{eq:MatrixAppHE}) that the coupling
matrices $A_{12}$ and $A_{21}$ become Hankel matrices. A Hankel
matrix have the following structure $A_{12}\left(m,\: n\right)=A_{12}\left(m-1,\: n+1\right)$,
i.e. the elements are identical when the sum of the row and column
indices are constant \cite{Peller2003}.

\subsection{Higher-dimensional systems}

For $D>1$, the summation operator that takes the system of creation
operators $\{\hat{a}_{\mathbf{k}',2}^{\dagger}\}_{\mathbf{k}'}$ to
the system of annihilation operators $\{\hat{a}_{\mathbf{k},1}\}_{\mathbf{k}}$
is defined via \begin{equation}
A_{12,\mathbf{n}}(\{\hat{a}_{\mathbf{n}',2}^{\dagger}\}_{\mathbf{n}'})=q\kappa\sum_{n_{1}'=-K}^{K}...\sum_{n_{D}'=-K}^{K}\tilde{g}_{n_{1}+n_{1}',...,n_{D}+n_{D}'}\hat{a}_{n_{1}',...,n_{D}',2}^{\dagger}.\label{eq:defHankelSuperOperator}\end{equation}
 We call this a $D$-dimensional finite Hankel operator. These have
been studied e.g. in \cite{HankelRef} for $D=2$. Obviously, there
are multiple ways to represent the two systems $\{\hat{a}_{\mathbf{k}',2}^{\dagger}\}_{\mathbf{k'}}$
and $\{\hat{a}_{\mathbf{k},1}\}_{\mathbf{k}}$ as a $2B^{D}-$dimensional
vector. However, with this done, the Hankel structure of the corresponding
coupling matrix $A_{12}$ is lost. In the next section we will construct
such a concrete representation in which $A_{12}$ turns out to be
a $D$-block-Hankel matrix.

\subsubsection{Ordering the lattice\label{sub:Ordering-the-lattice}}

Recall that $B=2K+1$. The following function \begin{equation}
f\left(n_{1},...,n_{D}\right)=1+\sum_{j=1}^{D}\left(n_{j}+K\right)B^{D-j},\label{HindexmapD}\end{equation}
 is a one-to-one mapping from the $D$-dimensional lattice $n_{1},...,n_{D},\,\: n_{j}\in\left\{ -K,\:...,\: K\right\} \,$
to $\,\left\{ 1,\:...,\: B^{D}\right\} $.

The inverse is

\begin{equation}
f^{-1}\left(m\right)=\left\{ \begin{array}{l}
n_{D}=mod\left(m-1,B\right)-K\\
\vdots\\
n_{d}=mod\left(m-1-\sum_{j=d+1}^{D}\left(n_{j}+K\right)B^{D-j},B^{D+1-d}\right)/B^{D-d}-K.\\
\vdots\\
n_{1}=\left(m-1-\sum_{j=2}^{D}\left(n_{j}+K\right)B^{D-j}\right)/B^{D-1}-K\end{array}\right.\label{HinversindexmapD}\end{equation}
 Note that the equations (\ref{HindexmapD}) and (\ref{HinversindexmapD})
are really nothing else than a change of base from $B$ to $10$ for
integer numbers.

We will later also use the following property of the map above

\begin{equation}
f^{-1}\left(B^{D}+1-m\right)=-f^{-1}\left(m\right).\label{eq:FirstPropertyOfMap}\end{equation}

\subsection{General construction of the system matrix\label{constrCoupling}}

We now identify the two lattices of operators $\{\widehat{a}_{\mathbf{n},1}\}$
and $\{\widehat{a}_{\mathbf{n},2}^{\dagger}\}$ with the corresponding
column vector $\left[\widehat{a}_{1,1}\ldots\widehat{a}_{B^{D},1}\widehat{a}_{1,2}^{\dagger}\ldots\widehat{a}_{B^{D},2}^{\dagger}\right]^{T}$,
where we use the simplified indices $1,\:...,\: B^{D}$ instead of
$f^{-1}(1),\:...,\: f^{-1}(B^{D})$. Then the $D$-dimensional system
(\ref{eq:MatrixAppHE_first_eq})-(\ref{eq:MatrixAppHE}) can be represented
in matrix form analogue to (\ref{ODEinMatrixForm}) as \begin{equation}
\frac{d}{dt}\left[\begin{array}{l}
\widehat{a}_{1,1}\\
\vdots\\
\widehat{a}_{B^{D},1}\\
\widehat{a}_{1,2}^{\dagger}\\
\vdots\\
\widehat{a}_{B^{D},2}^{\dagger}\end{array}\right]=\left[\begin{array}{cc}
A_{11} & A_{12}\\
A_{21} & A_{22}\end{array}\right]\left[\begin{array}{l}
\widehat{a}_{1,1}\\
\vdots\\
\widehat{a}_{B^{D},1}\\
\widehat{a}_{1,2}^{\dagger}\\
\vdots\\
\widehat{a}_{B^{D},2}^{\dagger}\end{array}\right].\label{ODEinMatrixForm1}\end{equation}
 Clearly, $A_{11}$ and $A_{22}$ are diagonal matrices whereas $A_{12}$
corresponds to the $D$-dimensional Hankel operator (\ref{eq:defHankelSuperOperator})
and $A_{21}$ to the lower counterpart in (\ref{eq:MatrixAppHE}).
As the matrices, $A_{12}$ and $A_{21}$ are not Hankel matrices for
$D>1$, but rather exhibit a block-Hankel structure \cite{HankelRef},
we will refer to them as $D$-block-Hankel matrices. In the coming
two sections we discuss properties of the four block matrices $A_{11},\: A_{12},\: A_{21}$
and $A_{22}$.

We will use the following notation; $T$ denotes transpose, $*$ is
complex conjugation, and $\dagger$ represents both the two previous
operations combined. Moreover, the operation of transposing in the
skew-diagonal will be denoted $SDT$, i.e. \begin{equation}
A^{SDT}(m^{R},m^{C})=A(B^{D}+1-m^{C},B^{D}+1-m^{R}),\label{eq54}\end{equation}
 or equivalently

\begin{equation}
A^{SDT}=\mathbb{S}A^{T}\mathbb{S},\label{eq:NEWeq54}\end{equation}
 where $\mathbb{S}$ denotes the skew-diagonal identity, i.e. the
matrix obtained by reversing the order of the columns of the identity
matrix $\mathbb{I}$. From (\ref{eq:NEWeq54}) and the property $\mathbb{S}\mathbb{S}=\mathbb{I}$,
it also follow that \textit{\emph{the skew-diagonal transpose of the
product of two general matrices $B_{1}$ and $B_{2}$ fulfills}}

\begin{equation}
(B_{1}B_{2})^{SDT}=B_{2}^{SDT}B_{1}^{SDT},\label{eq8}
\end{equation}
which will be used later. Finally, the skew-diagonal transpose combined
with complex conjugation will be denoted $SDH$.

\subsubsection{Structure of the $D$-block-Hankel matrices\label{sub:Structure-of-the-D-block-Hankel-matrices}}

Inspection of (\ref{eq:MatrixAppHE_first_eq})-(\ref{eq:MatrixAppHE})
and (\ref{ODEinMatrixForm1}) shows that $A_{12}$ is given elementwise
by \begin{equation}
A_{12}\left(m^{R},m^{C}\right)=q\kappa\tilde{g}_{f^{-1}\left(m^{R}\right)+f^{-1}\left(m^{C}\right)}.\label{eq45}\end{equation}
 If we denote the coordinates of $f^{-1}\left(m^{R}\right)$ by a
sup-index $R$, those of $f^{-1}\left(m^{C}\right)$ with a $C$ and
the coordinates for the Fourier coefficients by $n_{j}^{F}$, then
the coordinates for the rows and columns of $A_{12}$ fulfills

\begin{equation}
\left\{ \begin{array}{l}
n_{1}^{F}=n_{1}^{R}+n_{1}^{C}\\
\vdots\\
n_{D}^{F}=n_{D}^{R}+n_{D}^{C}\end{array}\right.,\label{HconstructA12for2D}\end{equation}
 which considerably simplify any practical implementation of $A_{12}$.
Note also that when $\tilde{g}_{\mathbf{k}}$ is defined on the same
$\mathbf{k}$-lattice as the atomic operators, we necessarily have
$\tilde{g}_{\mathbf{k}}\equiv0$ when $\left|n_{j}^{R}+n_{j}^{C}\right|>K$,
see figure \ref{fig:D-block_Hankel_matrix} for an illustration.

Similarly to (\ref{eq45}), we have \begin{equation}
A_{21}\left(m^{R},m^{C}\right)=\kappa\tilde{g}_{f^{-1}\left(m^{R}\right)+f^{-1}\left(m^{C}\right)}^{*},\label{eq:defA21Fromg}\end{equation}
 such that ($q^{2}=1$) \begin{equation}
A_{21}=qA_{12}^{*}.\label{eq:GeneralRelationForA21}\end{equation}
 Due to this identity we are satisfied with discussing properties
of $A_{12}$ in the remainder. First we note that \begin{equation}
A_{12}^{T}=A_{12},\label{eq:A12transposeA12}\end{equation}
 which is immediate by (\ref{eq45}).

To get additional structure, we impose extra conditions on the condensate
wave-function $\Psi=\sqrt{\rho\left(\mathbf{x}\right)}\exp\left(i\theta\left(\mathbf{x}\right)\right)$.
For many physical applications, $\Psi=\sqrt{\rho\left(\mathbf{x}\right)}$
can be chosen real. In this case, we note that the Fourier coefficients
(\ref{defFourierCoefficients}) have the following symmetry $\tilde{g}_{-\mathbf{k}}=\tilde{g}_{\mathbf{k}}^{*}$.
Therefore we get from~(\ref{eq:FirstPropertyOfMap}) \begin{equation}
A_{12}^{SDH}=A_{12},\label{eq:A12SDHequalA12}\end{equation}
 and from~(\ref{eq:GeneralRelationForA21}) that

\begin{equation}
A_{21}=qA_{12}^{SDT}.\label{eq:A21qA12SDT}\end{equation}
 Furthermore, for the physically important case of a condensate wave-function
that is real and even, i.e. with $\rho\left(\mathbf{x}\right)=\rho\left(-\mathbf{x}\right)$,
such as for example for a condensate in a harmonic trap, we have that
$\tilde{g}_{\mathbf{k}}$ is real so $A_{12}^{*}=A_{12}$, which combined
with (\ref{eq:A12SDHequalA12}) implies that \begin{equation}
A_{12}^{SDT}=A_{12},\label{eq:A12SDTequalA12}\end{equation}
 and from~(\ref{eq:GeneralRelationForA21}) that \begin{equation}
A_{21}=qA_{12}.\label{eq:EvenRelationForA21}\end{equation}

Finally, note that in the real uniform case, $\rho\left(\mathbf{x}\right)=\rho_{0},~\theta=0$,
we have non-zero entries only at positions where $\left(n_{1}^{R},...,n_{D}^{R}\right)=-\left(n_{1}^{C},...,n_{D}^{C}\right)$,
i.e. $A_{12}=qA_{21}=q\kappa\tilde{g}_{0}\mathbb{S}$, as in section~\ref{sub:Uniform--dimensional-MBEC}.

\subsubsection{Diagonal matrices for the kinetic energy}

According to the Heisenberg equations (\ref{eq:MatrixAppHE_first_eq})-(\ref{eq:MatrixAppHE})
and the representation (\ref{ODEinMatrixForm1}), the matrices $A_{11}$
and $A_{22}$ become diagonal (kinetic energy in momentum space),
where the $m$:th diagonal element of $A_{11}$ is given by \begin{equation}
A_{11}(m,m)=-i\Delta_{\mathbf{k}}=-i\left(\Omega+\frac{\hbar\left|\mathbf{k}\right|^{2}}{2m_{a}}\right),~\: m=f(\mathbf{k}),\label{eq:defA11}\end{equation}
 and similarly

\begin{equation}
A_{22}(m,m)=i\Delta_{\mathbf{k}}=A_{11}^{*}(m,m).\label{eq:A22equalA11*}
\end{equation}
Finally, we note that \begin{equation}
A_{11}^{SDT}=A_{11},\label{eq:A11SDTequalA11}\end{equation}
 which follows directly from (\ref{eq:FirstPropertyOfMap}), since
$|\mathbf{k}|=|-\mathbf{k}|$. 

\section{On the block-structure of the propagator\label{sub:A-Theorem-for}}

Much of the structure of the system-matrix $A$ is preserved in matrix
functions defined on $A$, which can be used to reduce the computational
complexity when evaluating $\exp\left(At\right)$ for the system's
evolution in time. Naturally, the more conditions we impose on the
condensate wave-function $\Psi$ in (\ref{defFourierCoefficients}),
the more structure is preserved. We present the corresponding identities
in order of increasing symmetry on $\Psi$, starting with a general
complex condensate wave-function. Throughout we will let $q=\pm1$
and prove the results for the cases of fermionic ($q=-1$) and bosonic
($q=1$) atoms simultaneously.

Given an arbitrary square (even sized) $2n\times2n$-matrix $B$ we
decompose it into its four blocks \[
B=\left[\begin{array}{cc}
B_{11} & B_{12}\\
B_{21} & B_{22}\end{array}\right].\]
 Below we list a number of useful matrix identities which all can
be verified by direct computation \begin{equation}
\left[\begin{array}{cc}
0 & q\mathbb{I}\\
\mathbb{I} & 0\end{array}\right]\left[\begin{array}{cc}
B_{11} & B_{12}\\
B_{21} & B_{22}\end{array}\right]\left[\begin{array}{cc}
0 & \mathbb{I}\\
q\mathbb{I} & 0\end{array}\right]=\left[\begin{array}{cc}
B_{22} & qB_{21}\\
qB_{12} & B_{11}\end{array}\right],\label{eq1}\end{equation}

\begin{equation}
\left[\begin{array}{cc}
B_{11} & B_{12}\\
B_{21} & B_{22}\end{array}\right]^{\dagger}=\left[\begin{array}{cc}
B_{11}^{\dagger} & B_{21}^{\dagger}\\
B_{12}^{\dagger} & B_{22}^{\dagger}\end{array}\right],\label{eq3}\end{equation}

\begin{equation}
\left[\begin{array}{cc}
B_{11} & B_{12}\\
B_{21} & B_{22}\end{array}\right]^{SDT}=\left[\begin{array}{cc}
B_{22}^{SDT} & B_{12}^{SDT}\\
B_{21}^{SDT} & B_{11}^{SDT}\end{array}\right].\label{eq4}\end{equation}
 With these three identities at hand, the following lemmas are all
immediate. For example the first one is a direct application of (\ref{eq1})
alone.

\subsection{Lemma}

\label{l1}\textit{ The matrix identities }

\textit{\begin{equation}
B_{11}=B_{22}^{*},~B_{12}=qB_{21}^{*},\label{eq6}\end{equation}
 are equivalent to \begin{equation}
B=\left[\begin{array}{cc}
0 & q\mathbb{I}\\
\mathbb{I} & 0\end{array}\right]B^{*}\left[\begin{array}{cc}
0 & \mathbb{I}\\
q\mathbb{I} & 0\end{array}\right].\label{eq6'}\end{equation}
 }

The second lemma follows by combining (\ref{eq1}) with (\ref{eq4}).

\subsection{Lemma}

\label{l2}\textit{ The matrix identities}

\textit{\begin{equation}
B_{11}=B_{11}^{SDT},~B_{22}=B_{22}^{SDT},~B_{12}=qB_{21}^{SDT},\label{eq5}\end{equation}
 are equivalent to \begin{equation}
B=\left[\begin{array}{cc}
0 & q\mathbb{I}\\
\mathbb{I} & 0\end{array}\right]B^{SDT}\left[\begin{array}{cc}
0 & \mathbb{I}\\
q\mathbb{I} & 0\end{array}\right].\label{eq5'}\end{equation}
 }

Finally combining (\ref{eq1}) with (\ref{eq3}) we obtain the third
lemma.

\subsection{Lemma}

\label{l3}\textit{ The matrix identities \begin{equation}
B_{11}=B_{22}^{\dagger},~B_{12}=B_{12}^{\dagger},~B_{21}=B_{21}^{\dagger},\label{eq7}\end{equation}
 are equivalent to \begin{equation}
B=\left[\begin{array}{cc}
0 & \mathbb{I}\\
\mathbb{I} & 0\end{array}\right]B^{\dagger}\left[\begin{array}{cc}
0 & \mathbb{I}\\
\mathbb{I} & 0\end{array}\right].\label{eq7'}\end{equation}
 }

\subsection{Proposition}

\label{p1} \textit{Let $B$ be the system-matrix $A$ constructed
in section \ref{constrCoupling} from the condensate wave-function
$\Psi$. Then the identities in (\ref{eq6}) of Lemma \ref{l1} are
always satisfied. Moreover, if $\Psi$ is real then the identities
in (\ref{eq5}) of Lemma \ref{l2} hold and if $\Psi$ is real and
even, the identities in (\ref{eq7}) of Lemma \ref{l3} hold.}

\textit{Proof.} The first claim follows by combining (\ref{eq:GeneralRelationForA21})
and (\ref{eq:A22equalA11*}), and the second follows by combining
(\ref{eq:A21qA12SDT}) and (\ref{eq:A11SDTequalA11}). When $\Psi$
is also even the elements of $A_{12}$ are real, and hence (\ref{eq:A12transposeA12})
can be written $A_{12}^{\dagger}=A_{12}$ (analogously $A_{21}^{\dagger}=A_{21}$).
Since clearly $A_{11}=A_{22}^{\dagger}$, the last claim is established
as well.

$\Box$

We need some preparation for the main theorem and corollaries, which
basically says that the above identities are preserved when forming
$\exp\left(At\right)$. First we note that \begin{equation}
\left[\begin{array}{cc}
0 & \mathbb{I}\\
q\mathbb{I} & 0\end{array}\right]\left[\begin{array}{cc}
0 & q\mathbb{I}\\
\mathbb{I} & 0\end{array}\right]=\left[\begin{array}{cc}
\mathbb{I} & 0\\
0 & q^{2}\mathbb{I}\end{array}\right]=\left[\begin{array}{cc}
\mathbb{I} & 0\\
0 & \mathbb{I}\end{array}\right].\label{eq2}\end{equation}

We now recall that an analytic function $\phi(z)=\sum_{k=0}^{\infty}c_{k}z^{k}$
defined on all of $\mathbb{C}$ is called an entire function, and
that for each convergence radius $r>0$ one can find a constant $C_{r}$
such that $|c_{k}|\leq\frac{C_{r}}{r^{k}}$, \cite{lang}. This allows
us to define the function $\phi(B)$ for any matrix $B$ (with the
corresponding matrix norm $\|B\|$), since \[
\|\sum_{k=N_{1}}^{N_{2}}c_{k}B^{k}\|\leq\sum_{k=N_{1}}^{N_{2}}|c_{k}|\|B^{k}\|\leq\sum_{k=N_{1}}^{N_{2}}C_{r}\frac{\|B\|^{k}}{r^{k}},\]
 which (picking any $r>\|B\|$) shows us that $(\sum_{k=0}^{N}c_{k}B^{k})_{N=0}^{\infty}$
is a Cauchy sequence in the set of matrices $\left\{ B\right\} $
with the operator norm $\|B\|$. The limit is thus a well defined
matrix since this is a Banach space. We recall that $\|B\|$ is defined
as $\|B\|=\sup\{\|B(x)\|:~\|x\|=1\}$ and in particular that for any
given index $(m,n)$ we have \begin{equation}
|B(m,n)|\leq\|B\|.\label{eq9}\end{equation}
 We are now ready for the main result.

\subsection{Theorem}

\label{t1} \textit{Let $\phi(z)=\sum_{k=0}^{\infty}c_{k}z^{k}$ be
an entire function and let $A$ be a matrix that satisfies either
of the identities in (\ref{eq6}), (\ref{eq5}) or (\ref{eq7}). Then
$\phi(A)$ satisfies the same identities.}

\textit{Proof.} Let us suppose that $A$ satisfies the identities
of (\ref{eq5}). By Lemma~\ref{l2} we then have (\ref{eq5'}), which
combined with (\ref{eq8}) and (\ref{eq2}), yields that \[
A^{k}=\left[\begin{array}{cc}
0 & q\mathbb{I}\\
\mathbb{I} & 0\end{array}\right](A^{k})^{SDT}\left[\begin{array}{cc}
0 & \mathbb{I}\\
q\mathbb{I} & 0\end{array}\right],\]
 for any $k$. Thus the matrix identities in (\ref{eq5}) are satisfied
for the corresponding blocks of $M=\phi(A)$ when $\phi$ is a monomial.
Since (\ref{eq5}) are also preserved upon taking linear combinations
of matrices that all satisfy (\ref{eq5}), we conclude that (\ref{eq5})
holds whenever $\phi$ is a polynomial. Finally, since in the general
case we have \[
\phi(A)=\lim_{N\rightarrow\infty}\sum_{k=0}^{N}c_{k}A^{k},\]
 in the operator norm, and since the identities (\ref{eq5}) are preserved
upon taking limits with respect to this norm {[}see (\ref{eq9}){]},
the general result follows. The proofs related to (\ref{eq6}) and
(\ref{eq7}) are analogue.

$\Box$

\textbf{Remark}: Connecting to Proposition \ref{p1}, we note that
when $B$ arise from a general complex condensate wave-function, one
still has more structure than what is expressed in (\ref{eq6}). For
example \[
-B_{11}=B_{22}^{T},~B_{12}=B_{12}^{T},~B_{21}=B_{21}^{T},\]
 which is easily seen to be equivalent to \[
B=\left[\begin{array}{cc}
0 & -\mathbb{I}\\
\mathbb{I} & 0\end{array}\right]B^{T}\left[\begin{array}{cc}
0 & -\mathbb{I}\\
\mathbb{I} & 0\end{array}\right].\]
 However, since \[
\left[\begin{array}{cc}
0 & -\mathbb{I}\\
\mathbb{I} & 0\end{array}\right]\left[\begin{array}{cc}
0 & -\mathbb{I}\\
\mathbb{I} & 0\end{array}\right]=\left[\begin{array}{cc}
-\mathbb{I} & 0\\
0 & -\mathbb{I}\end{array}\right],\]
 these properties are \textit{not} preserved in $\phi(B)$, except
for in the special case when $\phi$ is an odd function.

We now sum up our conclusions for the physically most interesting
cases.

\subsection{Corollary}

\label{c1} \textit{Given a real condensate wave-function $\Psi$
and $t\in\mathbb{R}$, set \begin{equation}
M=\exp\left(At\right)\equiv\sum_{j=0}^{\infty}\frac{\left(At\right)^{j}}{j!}.\label{eq:defExpAt}\end{equation}
 Then $M$ has the structure}

\begin{equation}
M=\left[\begin{array}{cc}
M_{11} & qM_{12}\\
M_{12}^{*} & M_{11}^{*}\end{array}\right],\label{eq:Corollary1RelationsForM}\end{equation}
 \textit{where in addition we have for the two blocks }

\begin{equation}
M_{11}^{SDT}=M_{11},\: M_{12}^{SDH}=M_{12}.\label{eq:Corollary1RelationsForBlocks}\end{equation}

\textit{Proof.} We can see that $M$ has the above structure if and
only if it satisfies the identities in (\ref{eq6}) and (\ref{eq5})
of Lemmas \ref{l1} and \ref{l2}. Moreover, $A$ satisfies these
identities by Proposition \ref{p1}. The desired conclusion thus follows
by Theorem~\ref{t1}.

$\Box$

It is clear from Corollary \ref{c1} that we only need to calculate
half of the matrices $M_{11}$ and $M_{12}$ in order to fully determine
$M$, which generally reduces the computational cost from $\left(2n\right)^{2}=4n^{2}$
to $2\left(n^{2}/2\right)=n^{2}$ elements in this case.

\subsection{Corollary}

\label{c2} \textit{Suppose that $\Psi$ is a real and even condensate
wave-function. Then, in addition to the identities in Corollary \ref{c1},
we have}

\begin{equation}
M_{11}^{T}=M_{11},\: M_{12}^{\dagger}=M_{12}.\label{eq:Corollary2RelationsForBlocks}\end{equation}

\textit{Proof.} By Proposition \ref{p1} $A$ satisfies the identities
in (\ref{eq7}) of Lemma \ref{l3}, and hence so does $M$ by Theorem
\ref{t1}. It is easy to see that these identities combined with the
structure in (\ref{eq:Corollary1RelationsForM}) and (\ref{eq:Corollary1RelationsForBlocks})
proven in Corollary \ref{c1} implies that the above identities are
satisfied for $M$ .

$\Box$

From Corollary \ref{c2} follows that we only need to calculate a
quarter of the matrices $M_{11}$ and $M_{12}$ in order to fully
determine $M$, which further reduces the computational cost to $2\left(n^{2}/4\right)=n^{2}/2$
elements in this case.

\section{Obtaining physical observables\label{sub:Obtaining-the-observables} }

In this section we show how to use the results of the previous section
in obtaining physical observables for the atoms. We start from the
following general block form of the solution $M=\exp\left(At\right)$
to Heisenberg equations (\ref{eq:MatrixAppHE_first_eq})-(\ref{eq:MatrixAppHE})
in matrix form (\ref{ODEinMatrixForm1})

\begin{equation}
\left[\begin{array}{l}
\widehat{a}_{1,1}\left(t\right)\\
\vdots\\
\widehat{a}_{B^{D},1}\left(t\right)\\
\widehat{a}_{1,2}^{\dagger}\left(t\right)\\
\vdots\\
\widehat{a}_{B^{D},2}^{\dagger}\left(t\right)\end{array}\right]=\left[\begin{array}{cc}
M_{11} & qM_{12}\\
M_{21} & M_{22}\end{array}\right]\left[\begin{array}{l}
\widehat{a}_{1,1}\left(0\right)\\
\vdots\\
\widehat{a}_{B^{D},1}\left(0\right)\\
\widehat{a}_{1,2}^{\dagger}\left(0\right)\\
\vdots\\
\widehat{a}_{B^{D},2}^{\dagger}\left(0\right)\end{array}\right],\label{MatrixSolutionToHeisenberg}\end{equation}
 It is obvious that the results of the previous section will simplify
the practical calculations of $M\left(t\right)$ and hence the physical
observables. However, first we show in the next section how to generally
obtain first-order moments for pairs of atomic operators directly
from (\ref{MatrixSolutionToHeisenberg}).

\subsection{First-order atomic moments\label{sub:First-order-atomic-moments}}

We now denote by $M_{ij,\mathbf{k}}$ the $m$-row-vector of the block
matrix $M_{ij}$, where $m=f\left(\mathbf{n}\right)$ is mapped to
$\frac{2\pi}{L}\mathbf{n}=\mathbf{k}=k_{1},...,k_{D}$ according to
section \ref{sub:Ordering-the-lattice}.

As a first example, for an annihilation operator of the $\sigma=1$
spin-state in row $m$ in the left hand side of (\ref{MatrixSolutionToHeisenberg})
we have

\begin{equation}
\widehat{a}_{m,1}\left(t\right)\equiv\widehat{a}_{\mathbf{k},1}\left(t\right)=M_{11,\mathbf{k}}\hat{u}+qM_{12,\mathbf{k}}\hat{v}=\hat{u}^{T}M_{11,\mathbf{k}}^{T}+q\hat{v}^{T}M_{12,\mathbf{k}}^{T},\label{eq:MatrixAppaf1Op}\end{equation}
 where we for notational and computational convenience introduce the
two operators 

\[
\hat{u}\equiv\left[\begin{array}{l}
\widehat{a}_{1,1}\left(0\right)\\
\vdots\\
\widehat{a}_{B^{D},1}\left(0\right)\end{array}\right],\:\:\hat{v}\equiv\left[\begin{array}{l}
\widehat{a}_{1,2}(0),\ldots,\widehat{a}_{B^{D},2}(0)\end{array}\right]^{\dagger}=\left[\begin{array}{l}
\widehat{a}_{1,2}^{\dagger}\left(0\right)\\
\vdots\\
\widehat{a}_{B^{D},2}^{\dagger}\left(0\right)\end{array}\right],\]
 which are naturally constructed from (\ref{MatrixSolutionToHeisenberg}).
With $\hat{u}$ and $\hat{v}$ we have introduced in effect a calculus,
where only terms in the expectation values containing the matrix $\left\langle \hat{u}\hat{u}^{\dagger}\right\rangle \equiv\mathbb{I}$
or $\left\langle \hat{v}^{\dagger^{T}}\hat{v}^{T}\right\rangle \equiv\mathbb{I}$
give a contribution, while all other terms are zero. This is due to
the (anti-) commutator relations applied to an initial vacuum state
for the atoms. See the examples leading to (\ref{eq:MatrixAppOOmkkprim})
and (\ref{eq:MatrixAppOOnkkprim1}) below for further details. With
these rules, different vacuum expectation values of atomic operator
pairs can conveniently be written down in terms of standard matrix
products between complex row- and column-vectors, where both the vectors
are defined from certain rows in one of the four block matrices $M_{ij}$.

We now give the corresponding expression to (\ref{eq:MatrixAppaf1Op})
for an annihilation operator of the $\sigma=2$ spin-state

\begin{equation}
\widehat{a}_{\mathbf{k},2}\left(t\right)=\left(\widehat{a}_{\mathbf{k},2}^{\dagger}\right)^{\dagger}=\left(M_{21,\mathbf{k}}\hat{u}+M_{22,\mathbf{k}}\hat{v}\right)^{\dagger}=\hat{u}^{\dagger}M_{21,\mathbf{k}}^{\dagger}+\hat{v}^{\dagger}M_{22,\mathbf{k}}^{\dagger}.\label{eq:MatrixAppOOak2}\end{equation}
 Note that analogously to (\ref{eq:MatrixAppaf1Op}), we can also
alternatively write (\ref{eq:MatrixAppOOak2}) on the form $\widehat{a}_{\mathbf{k},2}\left(t\right)=M_{21,\mathbf{k}}^{*}\hat{u}^{\dagger^{T}}+M_{22,\mathbf{k}}^{*}\hat{v}^{\dagger^{T}}$,
and for a specific moment one chose the option that allow the operators
to meet in the middle of the operator pair.

From (\ref{eq:MatrixAppaf1Op}) and (\ref{eq:MatrixAppOOak2}) we
can then demonstrate in detail how to calculate the anomalous moments,

\[
m_{\mathbf{k},\mathbf{k}'}\equiv\left\langle \widehat{a}_{\mathbf{k},1}\widehat{a}_{\mathbf{k}',2}\right\rangle =\left\langle \left(M_{11,\mathbf{k}}\hat{u}+qM_{12,\mathbf{k}}\hat{v}\right)\left(\hat{u}^{\dagger}M_{21,\mathbf{k}'}^{\dagger}+\hat{v}^{\dagger}M_{22,\mathbf{k}'}^{\dagger}\right)\right\rangle \]

\[
=\left\langle M_{11,\mathbf{k}}\hat{u}\hat{u}^{\dagger}M_{21,\mathbf{k}'}^{\dagger}+qM_{12,\mathbf{k}}\hat{v}\hat{u}^{\dagger}M_{21,\mathbf{k}'}^{\dagger}\right.\]

\[
\left.+M_{11,\mathbf{k}}\hat{u}\hat{v}^{\dagger}M_{22,\mathbf{k}'}^{\dagger}+qM_{12,\mathbf{k}}\hat{v}\hat{v}^{\dagger}M_{22,\mathbf{k}'}^{\dagger}\right\rangle \]

\[
=M_{11,\mathbf{k}}\left\langle \hat{u}\hat{u}^{\dagger}\right\rangle M_{21,\mathbf{k}'}^{\dagger}+qM_{12,\mathbf{k}}\left\langle \hat{v}\hat{u}^{\dagger}\right\rangle M_{21,\mathbf{k}'}^{\dagger}\]

\begin{equation}
+M_{11,\mathbf{k}}\left\langle \hat{u}\hat{v}^{\dagger}\right\rangle M_{22,\mathbf{k}'}^{\dagger}+qM_{12,\mathbf{k}}\left\langle \hat{v}\hat{v}^{\dagger}\right\rangle M_{22,\mathbf{k}'}^{\dagger}=M_{11,\mathbf{k}}M_{21,\mathbf{k}'}^{\dagger},\label{eq:MatrixAppOOmkkprim}\end{equation}
 which is a time-dependent complex number as expected.

In a similar way, using the Hermitian conjugate of (\ref{eq:MatrixAppaf1Op}),

\begin{equation}
\hat{a}_{\mathbf{k},1}^{\dagger}=\left(M_{11,\mathbf{k}}\hat{u}+qM_{12,\mathbf{k}}\hat{v}\right)^{\dagger}=\hat{u}^{\dagger}M_{11,\mathbf{k}}^{\dagger}+q\hat{v}^{\dagger}M_{12,\mathbf{k}}^{\dagger}=M_{11,\mathbf{k}}^{*}\hat{u}^{\dagger^{T}}+qM_{12,\mathbf{k}}^{*}\hat{v}^{\dagger^{T}},\label{eq:aDaggerK1}\end{equation}
 we have for the normal moments of the $\mbox{\ensuremath{\sigma}=1}$
spin-state

\[
n_{\mathbf{k},\mathbf{k}',1}\equiv\left\langle \hat{a}_{\mathbf{k},1}^{\dagger}\hat{a}_{\mathbf{k}',1}\right\rangle =\left\langle \left(M_{11,\mathbf{k}}^{*}\hat{u}^{\dagger^{T}}+qM_{12,\mathbf{k}}^{*}\hat{v}^{\dagger^{T}}\right)\left(\hat{u}^{T}M_{11,\mathbf{k}'}^{T}+q\hat{v}^{T}M_{12,\mathbf{k}'}^{T}\right)\right\rangle \]

\begin{equation}
=\left\langle qM_{12,\mathbf{k}}^{*}\hat{v}^{\dagger^{T}}\hat{v}^{T}qM_{12,\mathbf{k}'}^{T}\right\rangle =q^{2}M_{12,\mathbf{k}}^{*}M_{12,\mathbf{k}'}^{T}=M_{12,\mathbf{k}}^{*}M_{12,\mathbf{k}'}^{T}.\label{eq:MatrixAppOOnkkprim1}\end{equation}
 From (\ref{eq:MatrixAppOOak2}) it follows that the corresponding
result to (\ref{eq:MatrixAppOOnkkprim1}) for the $\mbox{\ensuremath{\sigma}=2}$
spin-state is

\[
n_{\mathbf{k},\mathbf{k}',2}\equiv\left\langle \hat{a}_{\mathbf{k},2}^{\dagger}\hat{a}_{\mathbf{k}',2}\right\rangle =\left\langle \left(M_{21,\mathbf{k}}\hat{u}+M_{22,\mathbf{k}}\hat{v}\right)\left(\hat{u}^{\dagger}M_{21,\mathbf{k}'}^{\dagger}+\hat{v}^{\dagger}M_{22,\mathbf{k}'}^{\dagger}\right)\right\rangle \]

\begin{equation}
=\left\langle M_{21,\mathbf{k}}\hat{u}\hat{u}^{\dagger}M_{21,\mathbf{k}'}^{\dagger}\right\rangle =M_{21,\mathbf{k}}M_{21,\mathbf{k}'}^{\dagger}.\label{eq:MatrixAppOOnkkprim2}\end{equation}

We finally confirm by direct calculations from (\ref{eq:MatrixAppOOak2})
and (\ref{eq:aDaggerK1}) that

\[
m_{\mathbf{k},\mathbf{k}'}^{*}\equiv\left\langle \left(\widehat{a}_{\mathbf{k},1}\widehat{a}_{\mathbf{k}',2}\right)^{\dagger}\right\rangle =\left\langle \widehat{a}_{\mathbf{k}',2}^{\dagger}\widehat{a}_{\mathbf{k},1}^{\dagger}\right\rangle =M_{21,\mathbf{k}'}M_{11,\mathbf{k}}^{\dagger}=\left(M_{11,\mathbf{k}}M_{21,\mathbf{k}'}^{\dagger}\right)^{*},\]
 as expected from (\ref{eq:MatrixAppOOmkkprim}).

Any other first-order moment that can be formed by two atomic operators,
such as for example $\left\langle \hat{a}_{\mathbf{k},1}^{\dagger}\hat{a}_{\mathbf{k}',2}\right\rangle $,
can be shown to be zero.

We now note that with $\mathbf{k}'=\mathbf{k}$ in (\ref{eq:MatrixAppOOnkkprim1})
we have

\[
n_{\mathbf{k},1}\equiv n_{\mathbf{k},\mathbf{k},1}=\left\langle \hat{a}_{\mathbf{k},1}^{\dagger}\hat{a}_{\mathbf{k},1}\right\rangle =M_{12,\mathbf{k}}^{*}M_{12,\mathbf{k}}^{T}=\sum\left|M_{12,\mathbf{k}}\right|^{2},\]
 which is a real number as expected from the physical interpretation
of the occupation of spin-$1$ atoms in state $\mathbf{k}$.

Motivated by (\ref{eq:MatrixAppOOnkkprim1}) and (\ref{eq:MatrixAppOOnkkprim2})
we see that the fact that the two spin-states are analogue in the
Fermi-Bose model (\ref{eq:HinSpatialCoordinates}), with initial atomic
vacuum in both spin-states, physically supports the identity $M_{21}=M_{12}^{*}$
given in Corollary \ref{c1}. Hence, applying Corollary \ref{c1}
it follows from (\ref{eq:MatrixAppOOnkkprim1}) and (\ref{eq:MatrixAppOOnkkprim2})
that we can write for arbitrary spin

\begin{equation}
n_{\mathbf{k},\mathbf{k}',\sigma}\equiv\left\langle \hat{a}_{\mathbf{k},\sigma}^{\dagger}\hat{a}_{\mathbf{k}',\sigma}\right\rangle =M_{12,\mathbf{k}}^{*}M_{12,\mathbf{k}'}^{T}.\label{eq:MatrixAppOOnkkprimsigma}\end{equation}
 Similarly the anomalous moments from (\ref{eq:MatrixAppOOmkkprim})
then take the form

\begin{equation}
m_{\mathbf{k},\mathbf{k}'}=M_{11,\mathbf{k}}M_{12,\mathbf{k}'}^{T},\label{eq:mkkprimFinal}\end{equation}
 in terms of only the blocks $M_{11}$ and $M_{12}$.

\subsection{Deduction of the uniform case\label{sub:Comparison-with-the}}

In this section we will deduce the solutions (\ref{eq:Uniform_nk})
and (\ref{eq:Uniform_mk}) direct from the exponentialmatrix in order
to check the formalism presented. We start by rewriting (\ref{eq:defExpAt})
according to

\textit{\begin{equation}
M=\exp\left(At\right)\equiv\sum_{j=0}^{\infty}\frac{\left(At\right)^{j}}{j!}=\sum_{j=0}^{\infty}\frac{A^{2j}t^{2j}}{\left(2j\right)!}+\sum_{j=0}^{\infty}\frac{A^{2j+1}t^{2j+1}}{\left(2j+1\right)!}.\label{eq:RewritingMin2sums}\end{equation}
 }\textit{\emph{Then starting from the system matrix for a uniform
molecular field with}} $g_{0}\equiv\kappa\tilde{g}_{0}=\chi\sqrt{\rho_{0}}$

\[
A=\left[\begin{array}{cc}
\textnormal{{diag}}\left(-i\vec{{\Delta_{\mathbf{k}}}}\right) & qg_{0}\mathbb{S}\\
g_{0}\mathbb{S} & \textnormal{{diag}}\left(i\vec{{\Delta_{\mathbf{k}}}}\right)\end{array}\right],\]
 we first note, using $\mathbb{S}^{2}=\mathbb{I}$, that the even
powers of $A$ are diagonal, hence we have

\begin{equation}
\sum_{j=0}^{\infty}\frac{\left(At\right)^{2j}}{\left(2j\right)!}=\left[\begin{array}{cc}
\textnormal{{diag}}\left(\cos\left(\sqrt{\vec{{\Delta_{\mathbf{k}}}}^{2}-qg_{0}^{2}}\, t\right)\right) & 0\\
0 & \textnormal{{diag}}\left(\cos\left(\sqrt{\vec{{\Delta_{\mathbf{k}}}}^{2}-qg_{0}^{2}}\, t\right)\right)\end{array}\right].\label{eq:FirstPartOfMSum}\end{equation}
 Secondly, we see by writing the odd powers as $AA^{2j}$ that

\[
\sum_{j=0}^{\infty}\frac{\left(At\right)^{2j+1}}{\left(2j+1\right)!}=\left[\begin{array}{l}
\textnormal{{diag}}\left(-i\frac{\vec{{\Delta_{\mathbf{k}}}}}{\sqrt{\vec{{\Delta_{\mathbf{k}}}}^{2}-qg_{0}^{2}}}\sin\left(\sqrt{\vec{{\Delta_{\mathbf{k}}}}^{2}-qg_{0}^{2}}\, t\right)\right)\\
\textnormal{{skew}}\left(\frac{g_{0}}{\sqrt{\vec{{\Delta_{\mathbf{k}}}}^{2}-qg_{0}^{2}}}\sin\left(\sqrt{\vec{{\Delta_{\mathbf{k}}}}^{2}-qg_{0}^{2}}\, t\right)\right)\end{array}\right.\]

\begin{equation}
\qquad\qquad\qquad\qquad\qquad\left.\begin{array}{r}
q\,\textnormal{{skew}}\left(\frac{g_{0}}{\sqrt{\vec{{\Delta_{\mathbf{k}}}}^{2}-qg_{0}^{2}}}\sin\left(\sqrt{\vec{{\Delta_{\mathbf{k}}}}^{2}-qg_{0}^{2}}\, t\right)\right)\\
\textnormal{{diag}}\left(i\frac{\vec{{\Delta_{\mathbf{k}}}}}{\sqrt{\vec{{\Delta_{\mathbf{k}}}}^{2}-qg_{0}^{2}}}\sin\left(\sqrt{\vec{{\Delta_{\mathbf{k}}}}^{2}-qg_{0}^{2}}\, t\right)\right)\end{array}\right].\label{eq:SecondPartOfMSum}\end{equation}
 Finally, we can from (\ref{eq:RewritingMin2sums}), (\ref{eq:FirstPartOfMSum})
and (\ref{eq:SecondPartOfMSum}) write up $M$ on the form of (\ref{MatrixSolutionToHeisenberg})

\begin{equation}
\left\{ \begin{array}{l}
M_{11}=M_{22}^{*}=\textnormal{{diag}}\left(\cos\left(\sqrt{\vec{{\Delta_{\mathbf{k}}}}^{2}-qg_{0}^{2}}\, t\right)-i\frac{\vec{{\Delta_{\mathbf{k}}}}}{\sqrt{\vec{{\Delta_{\mathbf{k}}}}^{2}-qg_{0}^{2}}}\sin\left(\sqrt{\vec{{\Delta_{\mathbf{k}}}}^{2}-qg_{0}^{2}}\, t\right)\right)\\
M_{12}=M_{21}^{*}=\textnormal{{skew}}\left(\frac{g_{0}}{\sqrt{\vec{{\Delta_{\mathbf{k}}}}^{2}-qg_{0}^{2}}}\sin\left(\sqrt{\vec{{\Delta_{\mathbf{k}}}}^{2}-qg_{0}^{2}}\, t\right)\right).\end{array}\right.\label{eq:AnalyticMforUniformCase}\end{equation}
 From (\ref{eq:AnalyticMforUniformCase}) we can then via (\ref{eq:MatrixAppOOnkkprimsigma})
and (\ref{eq:mkkprimFinal}) retrieve (\ref{eq:Uniform_nk}) and (\ref{eq:Uniform_mk})
of section \ref{sub:Uniform--dimensional-MBEC}.

However, it is important to stress that the (skew-) diagonal blockmatrices
of $M$ lose this structure for a non-uniform condensate wave-function,
and in the general case it follows by Cauchy-Schwarz that the equality
(\ref{eq:MatrixAppUniformSolutions}) changes into the following inequality

\begin{equation}
\left|m_{\mathbf{k}}\right|^{2}\leq n_{\mathbf{k},\sigma}\left(1+qn_{\mathbf{k},\sigma}\right).\label{eq:INEQUALITY_TO_PROVE-1}\end{equation}
 The above inequality have been investigated numerically for the case
of dissociation into bosonic atoms in \cite{SavagePRA2006} and it
implies limitations on the strength of various correlations for non-uniform
systems \cite{OgrenPRA2010,SavagePRA2006}.

\subsection{Higher-order atomic moments}

Higher-order moments, such as in the simplest case, the combination
of two pairs of operators, are factorized according to Wick's theorem
\cite{FetterWalecka}\textbf{ }which is implicit from the decorrelation
assumption in use within the undepleted molecular field approximation
here. As an example we calculate Glauber's correlation function for
two atoms in the same spin-state \cite{OgrenPRA2010}

\begin{equation}
g_{\sigma\sigma}^{\left(2\right)}\left(\mathbf{k},\mathbf{k}',t\right)\equiv\frac{\left\langle \widehat{a}_{\mathbf{k},\sigma}^{\dagger}\widehat{a}_{\mathbf{k}',\sigma}^{\dagger}\widehat{a}_{\mathbf{k}',\sigma}\widehat{a}_{\mathbf{k},\sigma}\right\rangle }{n_{\mathbf{k},\sigma}n_{\mathbf{k}',\sigma}}=1+\frac{q\left|n_{\mathbf{k},\mathbf{k}',\sigma}\right|^{2}}{n_{\mathbf{k},\mathbf{k},\sigma}n_{\mathbf{k}',\mathbf{k}',\sigma}}.\label{eq:MatrixAppOOg2kkprim}\end{equation}
 For a numerical implementation of (\ref{eq:MatrixAppOOg2kkprim})
we have, according to (\ref{eq:MatrixAppOOnkkprimsigma}),

\begin{equation}
g_{\sigma\sigma}^{\left(2\right)}\left(\mathbf{k},\mathbf{k}',t\right)=1+\frac{q\left|M_{12,\mathbf{k}}^{*}M_{12,\mathbf{k}'}^{T}\right|^{2}}{\left(M_{12,\mathbf{k}}^{*}M_{12,\mathbf{k}}^{T}\right)\left(M_{12,\mathbf{k}'}^{*}M_{12,\mathbf{k}'}^{T}\right)}.\label{eq:MatrixAppOOg2kkprimNUM}\end{equation}
 We here also give Glauber's correlation function for two atoms in
opposite spin-states

\begin{equation}
g_{12}^{\left(2\right)}\left(\mathbf{k},\mathbf{k}',t\right)\equiv\frac{\left\langle \widehat{a}_{\mathbf{k},1}^{\dagger}\widehat{a}_{\mathbf{k}',2}^{\dagger}\widehat{a}_{\mathbf{k}',2}\widehat{a}_{\mathbf{k},1}\right\rangle }{n_{\mathbf{k},1}n_{\mathbf{k}',2}}=1+\frac{\left|m_{\mathbf{k},\mathbf{k}'}\right|^{2}}{n_{\mathbf{k},\mathbf{k},1}n_{\mathbf{k}',\mathbf{k}',2}}.\label{eq:defBBcorr}\end{equation}
 For a numerical implementation of (\ref{eq:defBBcorr}) we have,
according to (\ref{eq:MatrixAppOOnkkprimsigma}) and (\ref{eq:mkkprimFinal}),

\begin{equation}
g_{12}^{\left(2\right)}\left(\mathbf{k},\mathbf{k}',t\right)=1+\frac{\left|M_{11,\mathbf{k}}M_{12,\mathbf{k}'}^{T}\right|^{2}}{\left(M_{12,\mathbf{k}}^{*}M_{12,\mathbf{k}}^{T}\right)\left(M_{12,\mathbf{k}'}^{*}M_{12,\mathbf{k}'}^{T}\right)}.\label{eq:BBcorrnum}\end{equation}

\subsection{Calculations of exponential matrices in practise\label{sec:Numerical-calculations-of}}

Up to this point, we have shown how to reduce the computational needs
for obtaining physical observables to the calculation of only a fraction
of the full exponential matrix $M=\exp\left(At\right)$. We now discuss
how we perform the necessary numerical calculations for the physical
observables in practice, while the results are presented in the next
section.

The matrix $\exp\left(At\right)$ can be calculated numerically in
many different ways, see for example \cite{Moler2003} for a review
of methods. For example, the two most obvious ones are: \emph{i)}
use the definition in (\ref{eq:defExpAt}) and calculate all powers
until some cut-off in $k$, $\left(At\right)^{k}/k!\approx0$, or
$=0$ if $A$ is nilpotent; \emph{ii)} diagonalize $A$ and use all
its eigenvalues ($\lambda_{j}$) and eigenvectors ($\mathbf{s}_{j}$)
to change to the $S$-basis, such that $M=SDS^{-1}$, where $D_{j,j}=\exp\left(\lambda_{j}t\right)$.
Both the natural methods \emph{i)} and \emph{ii)} in practice needs
many matrix-matrix operations, and have computationally expensive
performance for large matrices.

As explained in section \ref{sub:First-order-atomic-moments}, all
observables are obtained as matrix products between different row-
and column-vectors defined from rows of the blocks of $M$. It is
obvious that row $R$ of $M$ is obtained by multiplying $M^{T}$
with the unit vector $\mathbf{e}_{R}=\left[0,\:...,\:1_{R},\:...,\:0\right]^{T}$
from the right. For this task, powerful matrix-free algorithms exists
for general matrices \cite{Sidje1998}, i.e. that calculate the results
of an exponential matrix acting on an arbitrary vector, without the
need to perform any matrix-matrix operations.

\begin{figure}
\centering{}\includegraphics[scale=0.96]{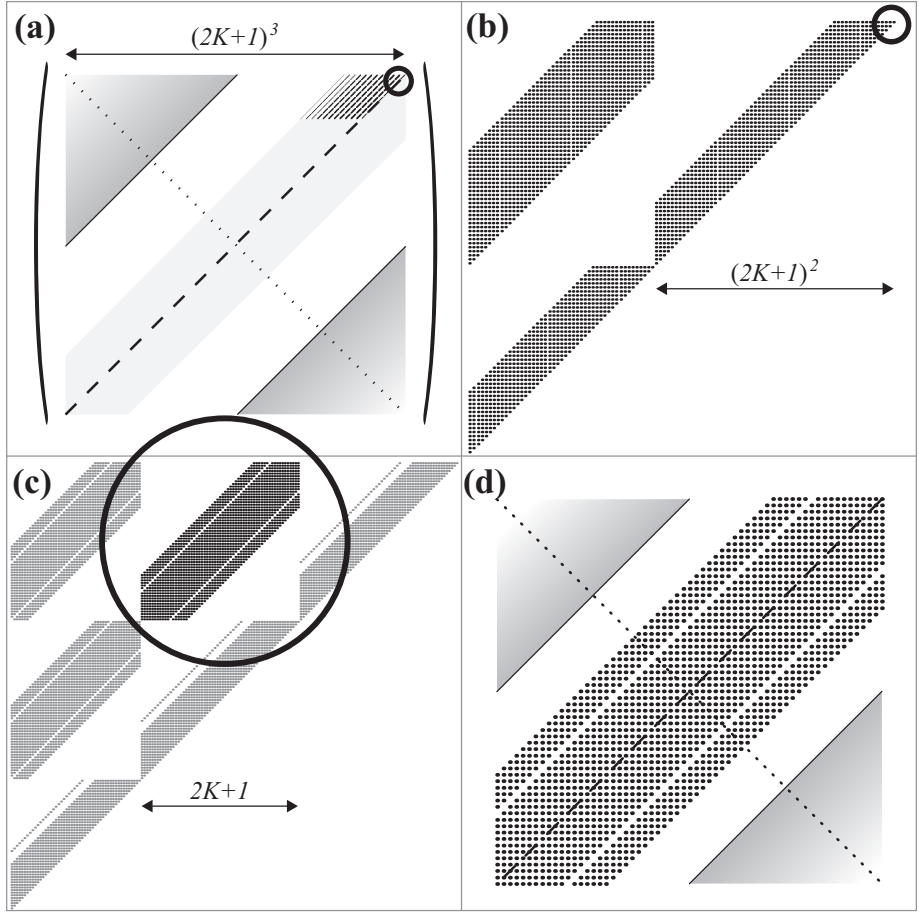}\caption{Illustration of the $D$-block-Hankel structure for the actual $A_{12}$
block-matrix used in the numerical example with $D=3$ and $K=30$
reported in this section. Figure (a) illustrates the entire $226981\times226981$
$A_{12}$ matrix, while (b), (c) and (d) shows zoomed in regions corresponding
to the circle of the previous subfigure. Note that in (b) each dot
represents a $2$-block-Hankel matrix, while in (c) and (d) each dot
represent a single non-zero matrix element. Dashed (dotted) lines
in (a) shows the skew-diagonal (diagonal) which corresponds to the
skew-diagonal transpose (SDT) (respectivelly SDH for complex matrix
elements) and transpose symmetries discussed in section~\ref{sub:Structure-of-the-D-block-Hankel-matrices}.
However, let us stress that the symmetry of a $D$-block-Hankel matrix
is considerable higher, since the block-matrices in (a) and (b) {[}elements
in (d){]} are repeated downwards parallell to the respective skew-diagonals.
Grey triangular regions in (a) and (d) shows where the conditions
$\left|n_{j}^{R}+n_{j}^{C}\right|>K$ for $j=3$ respectivelly $j=1$
are fulfilled. As discussed in section~\ref{sub:Structure-of-the-D-block-Hankel-matrices},
these regions does not contain any non-zero matrix elements due to
the restricted Fourier coefficients lattice size. The fact that the
regions of non-zero elements do not reach out to the grey fields are
due to the truncation implemented, see text for details. \label{fig:D-block_Hankel_matrix}}

\end{figure}

Using the results of the Corollaries in section \ref{sub:A-Theorem-for},
the fact that $A_{11}$ is diagonal and hence can be represented as
a vector, and the powerful property of $A_{12}$ being a $D$-block-Hankel
matrix, we use our own modified version of the open access software
Expokit for sparse matrices \cite{Sidje1998}. As will be reported
elsewhere, we have optimized the algorithm from \cite{Sidje1998}
for the implementation of $D$-block-Hankel matrices. This software
optimization, combined with the use of the results in section \ref{sub:A-Theorem-for},
give us the crucial advantages necessary for implementating large
Fermi-Bose systems, compared to any brute force calculation of the
full exponential matrix. While the complex matrix $M$, or any of
its blocks, is generally not sparse, the matrix $A_{12}$ is sparse
especially after truncation of the smallest Fourier coefficients,
see figure \ref{fig:D-block_Hankel_matrix} of the next session for
an example. Clearly any such actual truncation have to be evaluated
with convergence tests.

Overall, our optimized exponentiation procedure in effect reduces
the original problem of calculating the complex (non-sparse) $2n\times2n$-matrix
$M=\exp\left(At\right)$ to the simpler problem of performing matrix-vector
operations with the $D$-block-Hankel matrix $A_{12}^{T}=A_{12}$,
which in addition obey further symmetries, see e.g. (\ref{eq:A12transposeA12})
and (\ref{eq:A12SDHequalA12}). Furthermore, $A_{12}$ is real for
the physically important case of a condensate wave-function that is
even, which reduces the information to store in $A_{12}$ by an additional
factor of two in this case.

Finally, let us stress the important consequence of section \ref{sub:First-order-atomic-moments},
that the calculation of each row in the blocks $M_{ij}$ is independent,
such that the work can conviniently be distributed across several
computers in parallel to reduce computation time if needed. This is
in principle a very crucial advantage of the presented formalism,
when applied to large systems.

\section{Numerical illustrations\label{sec:Results}}

We here present specific numerical results for atomic correlation
functions for a non-isotropic three-dimensional system. We use a $61\times61\times61$
grid in momentum space, which corresponds to a system of linear differential
operator equations with a $2n\times2n$-system matrix where $n=61^{3}=226981$,
i.e. in general $\left(2n\right)^{2}\simeq2.1\cdot10^{11}$ matrix
elements. The theorem presented in section \ref{sub:A-Theorem-for}
in combination with the algorithm improvements briefly discussed in
section \ref{sec:Numerical-calculations-of} allow systems of this
size to be solved on one standard PC in the order of $\sim10$ hour.
As pointed out in section \ref{sec:Numerical-calculations-of}, substantially
larger grids can be attacked with calculations on several computers
in parallell.

\subsection{Physical parameters\label{sub:Parameters-of-the}}

In order to use a realistic set of parameters for our numerical example,
we chose an harmonic trap with frequences such that the size of the
molecular field along the $x$-direction $R_{TF,x}=8\,\mu$m is two
times the size along the $z$-direction $R_{TF,z}=4\,\mu$m, while
the size along the $y$-direction $R_{TF,y}=6\,\mu$m is set to an
intermediate value. With a central molecular peak density of $\rho_{0}=10^{20}$m$^{-3}$,
this corresponds to $N_{0}\simeq\frac{8\pi}{15}\rho_{0}R_{TF,x}R_{TF,y}R_{TF,z}\simeq3.2\cdot10^{4}$
molecules. Choosing $^{40}K_{2}$ dimers \cite{GreinerPRL2005} we
have an atomic mass of $m_{at}=6.642\cdot10^{-26}$kg. The molecule-atom
dissociation parameter is $\chi=10^{-7}$m$^{3/2}$/s here \cite{PMFT}.
We set the dissociation detuning to $\Omega=-4\cdot10^{3}$s$^{-1}$
which is large enough to ensure that the dissociation energy is larger
than thermal excitations at nK temperatures (i.e. $2\hbar\left|\Omega\right|\gg k_{B}T$).
With a characteristic time of $t_{0}=1$ms this corresponds to a dimensionless
detuning of $\delta=t_{0}\Omega=-4$ \cite{PMFT}. The momentum lattice
in use have a spacing $dk\equiv dk_{x}=dk_{y}=dk_{z}\simeq1.1\cdot10^{5}$m$^{-1}$
which is smaller than the smallest width of the molecular momentum
distribution $\sim2/R_{TF,x}$ \cite{OgrenPRA2010}, and have been
confirmed numerically to resolve the dynamics of the relevant structures
in the atomic momentum distribution. The corresponding resonance momenta
is then $k_{0}=\left|\mathbf{k}_{0}\right|=\sqrt{2m_{at}\left|\Omega\right|/\hbar}\simeq20dk$.

\subsubsection{Fermi's Golden Rule estimate of the atom numbers}

For small times we can, for the purpose of validating the physical
parameters, estimate the number of atoms by the following linear expression
in time

\begin{equation}
N_{j}\left(t\right)\simeq N_{0}\lambda t,\:\:\lambda=\frac{1}{\sqrt{2}\pi}\left(\frac{m_{at}}{\hbar}\right)^{3/2}\chi^{2}\sqrt{\left|\Omega\right|}.\label{eq:FERMI-GOLDEN-RULE}\end{equation}
 From the above formula it is clear that the number of atoms increase
with $\left|\Omega\right|$ for a three-dimensional system. This have
earlier been studied explicitly for a uniform system, see Fig.~1
of \cite{PMFT}. Hence for cases with large detuning the validity
of the undepleted field approximation is limited to short dissociation
times $t/t_{0}\ll1$. For the parameters of section \ref{sub:Parameters-of-the}
we have $\lambda\sim2s^{-1}$ which results in $N_{j}\sim10^{2}$
atoms at $t=t_{0}$, and hence a conversion ratio of less than 1\%,
ensuring the validity of the results from the undepleted field approximation
\cite{OgrenPRA2010,MidgleyPRA2009,Magnus-Karen-Joel-First-Principles}.
The presented estimate of atom numbers from the Fermi's Golden Rule
(\ref{eq:FERMI-GOLDEN-RULE}) was later confirmed by the numerical
calculations for the parameter values in use here.

\begin{figure}
\includegraphics[scale=0.31]{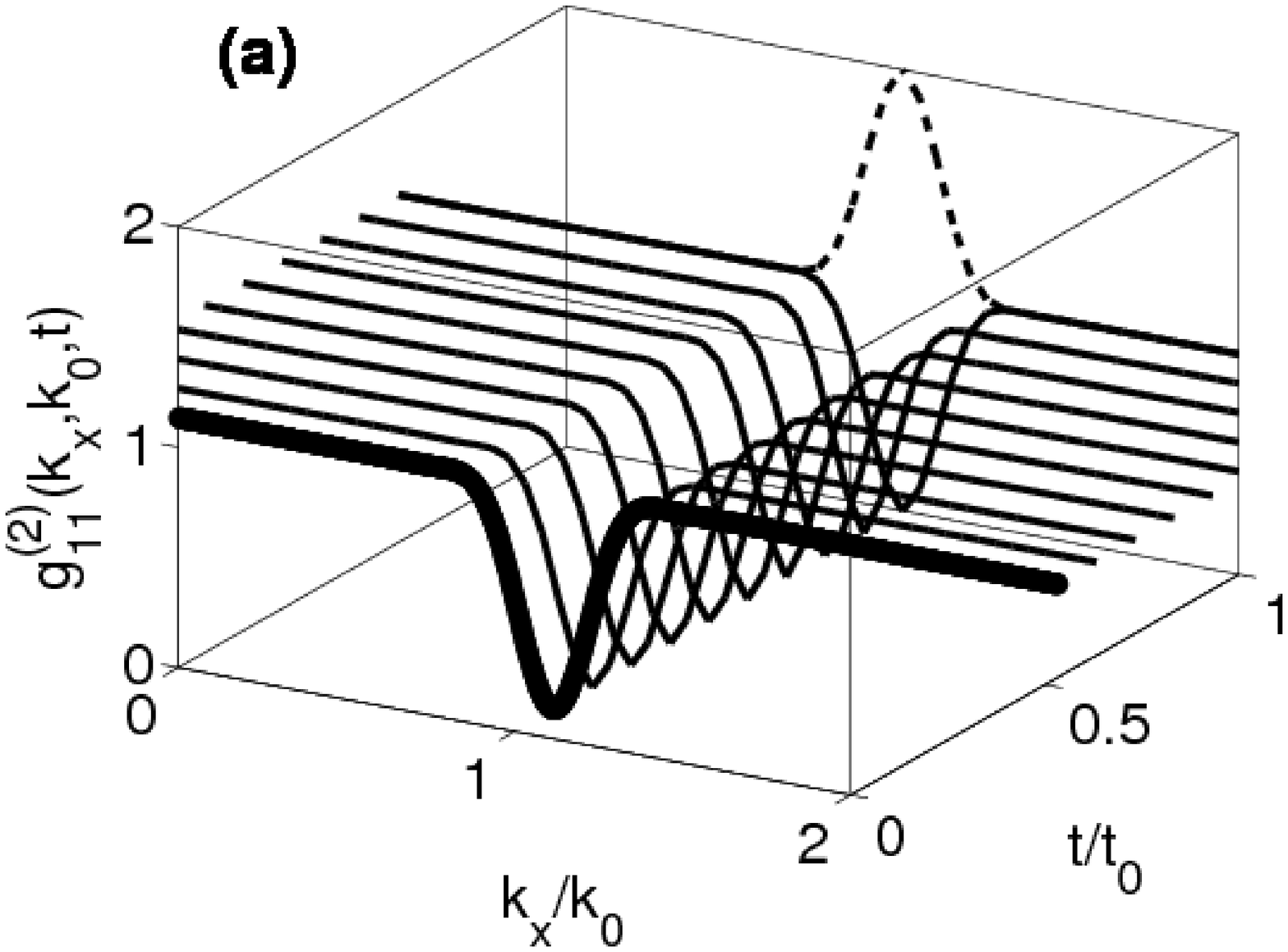}\includegraphics[scale=0.31]{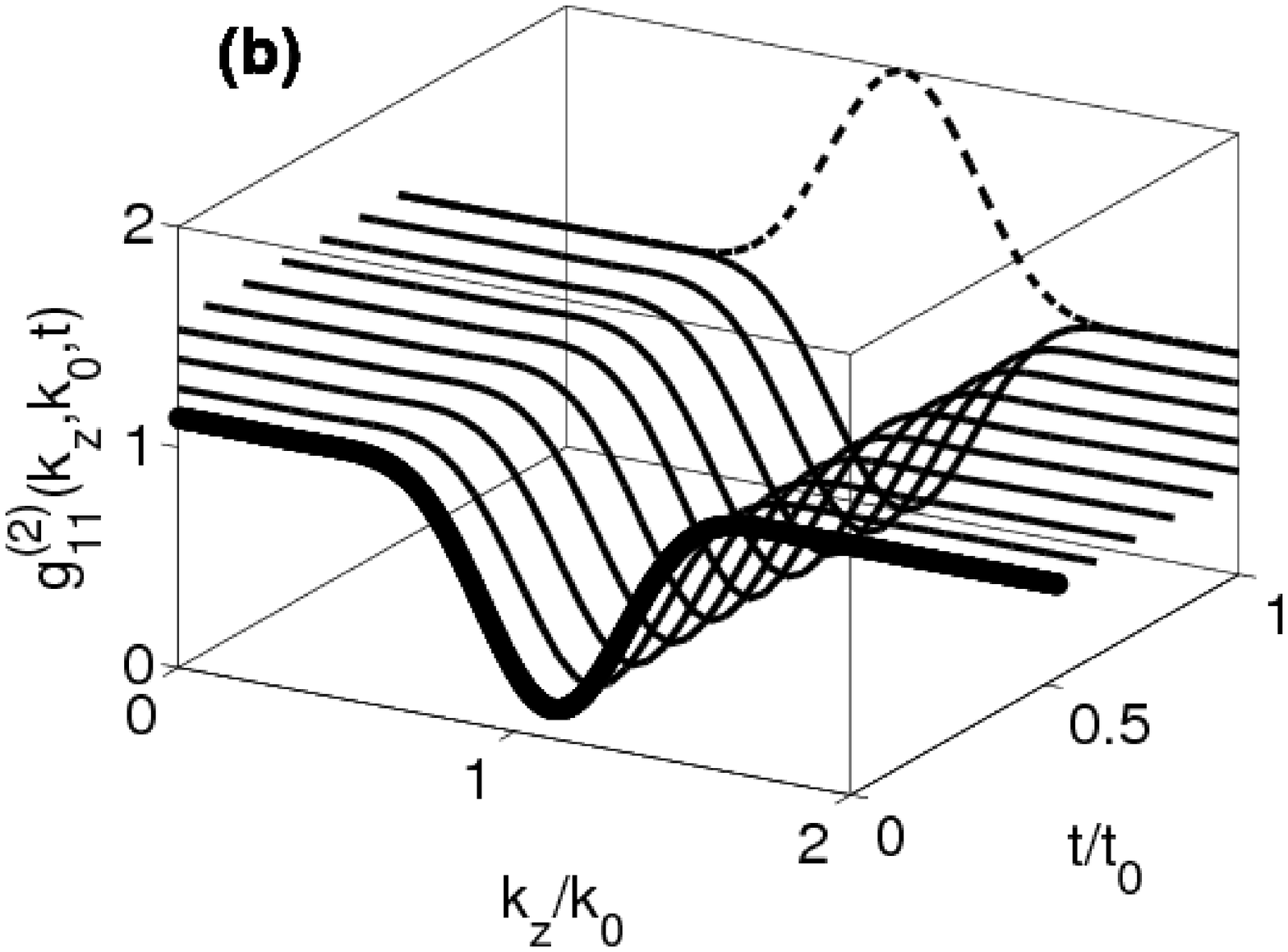}\caption{Fermionic collinear atom-atom correlation functions in momentum space
$g_{11}^{(2)}(\mathbf{k},\mathbf{k}',t)$, at times $t/t_{0}=0.1,\:0.2,\:...,\:1$
($t_{0}=1$ms), calculated along different directions in 3D. In (a)
we show numerical results (solid thin curves) from (\ref{eq:MatrixAppOOg2kkprimNUM})
along the direction~$\mathbf{e}_{x}$, i.e. with $\mathbf{k}=k_{x}\mathbf{e}_{x}$
and $\mathbf{k}'=k_{0}\mathbf{e}_{x}$ ($k_{0}\simeq2.2\cdot10^{6}$m$^{-1}$),
while in (b) we show the corresponding result along the direction~$\mathbf{e}_{z}$.
The analytic short-time asymptotes of (\ref{eq:g_jj^2_ana}) are represented
by the fat curves plotted at $t/t_{0}=0.1$ only. In fact the short-time
asymptotes for the collinear correlations are in qualitative agreement
with the numerical results even up to $t/t_{0}\sim1$. However, a
zoom in reveals quantitative deviations seen as a narrowing of the
width of the correlation signal with time, this is also in agreement
with detailed 1D results reported in Fig.~8 of~\cite{OgrenPRA2010}.
In general the fermionic collinear correlations are here showing a
Pauli-blocking dip at $k_{x,z}=k_{0}$, while the characteristic width
of the correlation signal have been confirmed to be inverselly proportional
to the size of the molecular BEC source along the corresponding direction,
i.e. $\sim2.16R_{TF,x}^{-1}\simeq2.7\cdot10^{5}$m$^{-1}$ in (a)
and $\sim2.16R_{TF,z}^{-1}=2\cdot2.16R_{TF,x}^{-1}$ in (b). In addition
we compared with the corresponding results for bosonic atoms showing
a so called Hanbury-Brown and Twiss peak at $k_{x,z}=k_{0}$ (dashed
curves), shown only for the largest time here.\label{fig:Atom-atom-correlation-functions}}

\end{figure}

\subsection{Structures in the system matrix}

We here explicitly illustrate the $D$-block-Hankel matrix $A_{12}$
that is used in the numerical calculations of a physical system for
$D=3$ here. Hence, it is evident from the figures~\ref{fig:D-block_Hankel_matrix}~(a),~(b)
and~(c) that we can zoom in $D=3$ times on $A_{12}$ and reveal
a repeating pattern. Clearly after the last zoom in {[}figure~\ref{fig:D-block_Hankel_matrix}~(d){]},
we are left with a structure of a usual Hankel matrix.

We have used a truncation of the molecular BEC source such that Fourier
coefficients with a modulus less than 2\% of the leading coefficient
is neglected. This procedure have been evaluated by reconstruction
of the BEC by the inverse Fourier transform. It was also found that
the 2\% level of truncation resulted in correlation functions (see
figure~\ref{fig:Atom-atom-correlation-functions}) that could not
be distingushable by the eye from the correlation functions obtained
with a 4\% level of truncation.

\subsection{Numerical comparison with analytic asymptotes}

For $D=3$ the Thomas-Fermi (TF) density profile of the molcular BEC
is given by $\rho_{0}(\mathbf{x})=\rho_{0}(1-x^{2}/R_{\mathrm{TF},x}^{2}-y^{2}/R_{\mathrm{TF},y}^{2}-z^{2}/R_{\mathrm{TF},z}^{2})$
for $x^{2}/R_{\mathrm{TF},x}^{2}+y^{2}/R_{\mathrm{TF},y}^{2}+z^{2}/R_{\mathrm{TF},z}^{2}<1$
{[}and $\rho_{0}(\mathbf{x})=0$ otherwise{]}, which is underlying
an analytic derivation of the asymptotes. Here $R_{\mathrm{TF},j}$
is the Thomas-Fermi radius along the spatial direction $j=x,y,z$.
We are here interested in collinear (CL) density correlations between
two momentum components at $\mathbf{k}$ and $\mathbf{k}^{\prime}$,
for which the displacement $\Delta\mathbf{k=k-k}^{\prime}$ is along
one of the Cartesian coordinates, $k_{j}$. The detailed derivation
of short-time asymptotes for the correlation functions in this case
was reported in \cite{OgrenPRA2010}. The CL correlations following
from this derivation is \begin{equation}
g_{11}^{(2)}(k_{j},k_{j}^{\prime},t)\simeq1+q\frac{225\pi}{2}\frac{\left[J_{5/2}\left((k_{j}-k_{j}^{\prime})R_{\mathrm{TF},j}\right)\right]^{2}}{\left[(k_{j}-k_{j}^{\prime})R_{\mathrm{TF},j}\right]^{5}},\label{eq:g_jj^2_ana}\end{equation}
 where $J_{\nu}$ denotes Bessel functions of the first kind. The
qualitative behavior of the CL correlation functions are similar as
in lower dimensions \cite{Ogren-directionality,Magnus-Karen-dissociation-PRA-Rapid},
whereas the quantitative differences enter e.g. through the width
and the peak values. The widths of (\ref{eq:g_jj^2_ana}) is $w_{i}^{\left(CL\right)}\simeq2.16/R_{TF,j}$
and the peak value of (\ref{eq:g_jj^2_ana}) is static to leading
order \cite{OgrenPRA2010}.

In figure \ref{fig:Atom-atom-correlation-functions} we show results
for the evaluation of the analytic short-time asymptote (\ref{eq:g_jj^2_ana}),
strictly valid in the $t/t_{0}\ll1$ limit, against numerical results
for times $t/t_{0}\leq1$, where $t=t_{0}$ roughly corresponds to
the first maximum in time of the oscillating fermionic atom numbers
$N\left(t\right)=\sum_{\mathbf{k},\sigma}n_{\mathbf{k},\sigma}\left(t\right)$.

\section{Summary\label{sec:Summary}}

We have described how to effectivelly calculate the dynamics of linear
Heisenberg operator equations for the Fermi-Bose model applied to
the problem of molecular dissociation. We note that a similar framework
have been used to obtain numerical results for a non-isotropic 2D
system on a $61\times61$ grid in \cite{Ogren-directionality}. We
have here generalized the approch to $D$ spatial dimensions with
the use of $D$-block-Hankel matrices. In particular we have explicitly
explored a non-isotropic 3D system on a $61\times61\times61$ grid
numerically on a standard PC. Such a grid can resolve relevant atom
dynamics in momentum space for realistic parameters \cite{Fermidiss},
and naturally extends earlier studies of non-uniform 1D and 2D systems
\cite{Ogren-directionality,Magnus-Karen-dissociation-PRA-Rapid},
and is more realistic than previous treatments of uniform 3D systems
\cite{Fermidiss,PMFT}. We finally stress that the results presented
can be used to handle a complex bosonic mean-field of any geometry
in any spatial dimension.

\section*{Acknowledgments}

We thank Kar\'{e}n Kheruntsyan and Roger Sidje for valuable discussions
at an early stage, and Johnny Kvistholm for artistic assistance with
figure~1.

\end{document}